\documentclass[apjl]{emulateapj}

\usepackage{graphicx}
\usepackage{t1enc}
\usepackage[varg]{txfonts}
\usepackage{epsfig}
\usepackage{rotating}
\usepackage{natbib}
\usepackage{color}
\newcommand{\Nmm} 	{$N_\mathrm{mm}$} 
\newcommand{\Lbol}		{$L_\mathrm{bol}$}
\newcommand{\Tbol}  	{$T_\mathrm{bol}$}

\newcommand{\mum}   	{$\mu$m}
\newcommand{\kms}   {km~s$^{-1}$}

\newcommand{\cmt}   {cm$^{-3}$}
\newcommand{\jpb}   {$\rm Jy~beam^{-1}$}    
\newcommand{\lo}    {$L_{\sun}$}
\newcommand{\mo}    {$M_{\sun}$}
\newcommand{\nh}    {NH$_3$}

\newcommand{\chtoh}  {CH$_3$OH}

\newcommand{\et}    {et al.}
\newcommand{\eg}    {e.\,g.,}
\newcommand{\ie}    {i.\,e.,}

\newcommand{\supa}  {$^\mathrm{a}$}
\newcommand{\supb}  {$^\mathrm{b}$}
\newcommand{\supc}  {$^\mathrm{c}$}
\newcommand{\supd}  {$^\mathrm{d}$}
\newcommand{\supe}  {$^\mathrm{e}$}
\newcommand{\supf}  {$^\mathrm{f}$}
\newcommand{\supg}  {$^\mathrm{g}$}
\newcommand{\suph}  {$^\mathrm{h}$}
\newcommand{\supi}  {$^\mathrm{i}$}

\definecolor{RED}{rgb}{1.0,0.0,0.0}
\shorttitle{Fragmentation of massive cores down to $\sim1000$~AU}
\shortauthors{Palau et al.}

\begin{document}

\title{Early stages of cluster formation:\\
fragmentation of massive dense cores down to $\lesssim1000$~AU\footnote{Based on observations carried out with the IRAM Plateau de Bure Interferometer. IRAM is supported by INSU/CNRS (France), MPG (Germany) and IGN (Spain).}
}

\author{Aina Palau\altaffilmark{1},
Asunci\'on Fuente\altaffilmark{2},
Josep M. Girart\altaffilmark{1},
Robert Estalella\altaffilmark{3},
Paul T. P. Ho\altaffilmark{4,5},
\'Alvaro S\'anchez-Monge\altaffilmark{6},
Francesco Fontani\altaffilmark{6},
Gemma Busquet\altaffilmark{7},
Benoit Commer\c con\altaffilmark{8},
Patrick Hennebelle\altaffilmark{8},
J\'er\'emie Boissier\altaffilmark{9,10},
Qizhou Zhang\altaffilmark{4},
Riccardo Cesaroni\altaffilmark{6},
Luis A. Zapata\altaffilmark{11}
}
\altaffiltext{1}{Institut de Ci\`encies de l'Espai (CSIC-IEEC), Campus UAB -- Facultat de Ci\`encies, Torre C5 -- parell 2, 08193 Bellaterra, Catalunya, Spain}
\email{palau@ieec.uab.es}
\altaffiltext{2}{Observatorio Astron\'omico Nacional, P.O. Box 112, 28803 Alcal\'a de Henares, Madrid, Spain}
\altaffiltext{3}{Departament d'Astronomia i Meteorologia (IEEC-UB), Institut de Ci\`encies del Cosmos, Universitat de Barcelona, Mart\'i Franqu\`es, 1, 08028 Barcelona, Spain}
\altaffiltext{4}{Harvard-Smithsonian Center for Astrophysics, 60 Garden Street, Cambridge, MA 02138, USA}
\altaffiltext{5}{Institute of Astronomy and Astrophysics, Academia Sinica, P.O. Box 23-141, Taipei 106, Taiwan}
\altaffiltext{6}{Osservatorio Astrofisico di Arcetri, INAF, Lago E. Fermi 5, 50125, Firenze, Italy}
\altaffiltext{7}{INAF-Istituto di Astrofisica e Planetologia Spaziali, Area di Recerca di Tor Vergata, Via Fosso Cavaliere 100, 00133, Roma, Italy}
\altaffiltext{8}{Laboratoire de Radioastronomie, UMR CNRS 8112, \'Ecole Normale Sup\'erieure et Observatoire de Paris, 24 rue Lhomond, 75231 Paris Cedex 05, France} 
\altaffiltext{9}{Istituto di Radioastronomia, INAF, Via Gobetti 101, Bologna, Italy}
\altaffiltext{10}{ESO, Karl Schwarzschild St. 2, 85748 Garching Muenchen, Germany}
\altaffiltext{11}{Centro de Radioastronom\'ia y Astrof\'isica, Universidad Nacional Aut\'onoma de M\'exico, P.O. Box 3-72, 58090, Morelia, Michoac\'an, Mexico}

\begin{abstract}
In order to study the fragmentation of massive dense cores, which constitute the cluster cradles, we observed with the PdBI in the most extended configuration the continuum at 1.3~mm and the CO\,(2--1) emission of four massive cores. We detect dust condensations down to $\sim0.3$~M$_\odot$ and separate millimeter sources down to $0.4''$ or $\lesssim1000$~AU, comparable to the sensitivities and separations reached in optical/infrared studies of clusters. The CO\,(2--1) high angular resolution images reveal high-velocity knots usually aligned with previously known outflow directions. This, in combination with additional cores from the literature observed at similar mass sensitivity and spatial resolution, allowed us to build a sample of 18 protoclusters with luminosities spanning 3 orders of magnitude. Among the 18 regions,  $\sim30$\% show no signs of fragmentation, while $50$\% split up into $\gtrsim4$ millimeter sources. We compiled a list of properties for the 18 massive dense cores, such as bolometric luminosity, total mass, and mean density, and found no correlation of any of these parameters with the fragmentation level. In order to investigate the combined effects of magnetic field, radiative feedback and turbulence in the fragmentation process, we compared our observations to radiation magneto-hydrodynamic simulations, and obtained that the low-fragmented regions are well reproduced in the magnetized core case, while the highly-fragmented regions are consistent with cores where turbulence dominates over the magnetic field. Overall, our study suggests that the fragmentation in massive dense cores could be determined by the initial magnetic field/turbulence balance in each particular core.
\end{abstract}


%%%%%%%%%%%%%%%%%%%%%%%%%%%%%%%%%%%%%%%%%%%%%%%%%%%%%%%%%%%%%%%%%%%%%%%%%%%%%%%%
\section{Introduction \label{si}}

A well-observed trend is that most stars form in clusters/groups,
rather than in isolation (\eg\ Carpenter 2000; Lada \& Lada 2003), and
thus the understanding of cluster formation is intimately related to
the understanding of star formation. One of the results found from statistical studies of clusters at optical/infrared wavelengths, carried out down to mass sensitivities of 0.1--0.5~\mo\ and spatial resolutions of $\sim1000$~AU,  is that there is a correlation between the mass of the most
massive star in the cluster and the number of members of the cluster
(\eg\ Hillenbrand 1995; Testi \et\ 1999; Qiu \et\ 2008). However, it is not clear which is
the physical reason for this correlation: it could either be the
result of the initial mass function (\eg\ Testi \et\ 2000; Parker \& Goodwin 2007), or it
could indicate that the most massive stars need a rich cluster to be
formed (\eg\ Stahler \et\ 2000; Krumholz \& Bonnell 2007). Alternatively, the most massive
stars may play an important role to increase the number of members in
their surroundings, by inducing the formation of new stars (\eg\ Hester \& Desch 2005; Deharveng \et\ 2012). 
Thus, it is necessary to investigate whether the mentioned correlation is already established at the beginning of the formation of the cluster, or it is a result of cluster evolution.

To answer these questions about cluster formation it is of fundamental importance to characterize clusters in their very first stages of formation, when they are deeply embedded in a massive dense core and thus detectable in the millimeter/submillimeter range. However, 
in general the present millimeter/submillimeter instrumentation does not allow to reach the required mass sensitivity and spatial resolution ($\lesssim0.5$~\mo, $\sim1000$~AU),
given the typical distances of massive dense cores of $>1$~kpc, in a reasonable amount of time for a sample of regions.  
Thus, several studies have focused on millimeter observations of \emph{individual} massive cores, associated with intermediate/high-mass protostars, down to spatial scales of $\lesssim 1000$~AU and reaching mass sensitivities around 0.1~\mo\ (\eg\ Beuther et al. 2004; Zapata et al. 2005; Fuente et al. 2005; Leurini et al. 2007; Neri et al. 2007; Teixeira et al. 2007; Rod\'on et al. 2008; Su et al. 2009; Tang et al. 2010), but no studies up to date focus on the formation of clusters from a statistical point of view. 
More recently, massive cores at even earlier evolutionary stages (being dark or `quiet' in the infrared --the so-called massive infrared dark clouds), have been imaged at high angular resolution to investigate fragmentation (\eg\ Swift 2009; Zhang \et\ 2009; Bontemps \et\ 2010; Pillai \et\ 2011; Zhang \& Wang 2011; Wang \et\ 2011; Wang \et\ 2012b). In particular, Bontemps et al. (2010) imaged with the Plateau de Bure Interferometer (PdBI) in compact (CD) configuration six IR-quiet massive cores in Cygnus X, studying fragmentation down to 1400~AU, and find that the fragmentation level is higher than expected for a turbulence regulated monolithic collapse scenario (\eg\ McKee \& Tan 2002, 2003), but lower than expected for a pure gravo-turbulent scenario (\eg\ Klessen \et\ 2005; Bonnell \et\ 2007; see Csengeri \et\ 2011 for further details on both scenarios). This calls for further studies to understand the fragmentation process.
In this paper we report on observations carried out with the PdBI operated in A configuration toward four massive dense cores associated with intermediate/high-mass protostars. These PdBI millimeter observations provide a spatial resolution and sensitivity comparable to those reached in the optical/infrared studies. These four regions, completed with the regions from the literature mentioned above, 
constitute a step towards a statistical study of the first stages of cluster formation.

The four observed regions were selected from the literature following these criteria: 
i) regions must have been observed with a millimeter/submillimeter interferometer in a compact configuration (with angular resolution $\lesssim7''$), and must show a compact condensation of gas and/or dust, to assure that the intermediate/high-mass protostar is deeply embedded in a massive dense core; 
ii) regions must be at a distance $\lesssim3$~kpc, to obtain the required high spatial resolution and high mass sensitivity; 
iii) regions must span a range of luminosities between $\sim300$~\lo\ and $\sim10^5$~\lo, which is the range in which clustering becomes dominant in the star formation process; 
iv) regions must belong to different cloud complexes to avoid biases due to similar initial conditions; and
v) regions had to be observable from the PdBI when the A configuration was operative. 
These selection criteria will allow us to study the richness of the protoclusters as function of the luminosity and compare it to the optical/infrared results found for more evolved clusters.
The four selected regions are IRAS\,22198+6336 (I22198); a region $\sim15''$ north of IRAS\,22172+5549 (I22172N); AFGL\,5142 (A5142); and IRAS\,22134+5834 (I22134). I22198 is a Class 0 source of $\sim300$~\lo\ harboring a hot core and driving a quadrupolar outflow (S\'anchez-Monge et al. 2010). I22172N and A5142 have luminosities of $\sim800$ and 2200~\lo, respectively, are also classified as Class 0 sources, and are associated with outflows (\eg\ Hunter \et\ 1999; Fontani \et\ 2004; Zhang \et\ 2007). Finally, I22134, with a luminosity of $\sim12000$~\lo, is an UCHII region embedded within a dust condensation (S\'anchez-Monge 2011). 
Thus, all the selected regions harbor intermediate/high-mass protostars deeply embedded in massive dense cores, being excellent protocluster candidates. 
We will follow the nomenclature of Williams \et\ (2000) and Bontemps \et\ (2010) where a massive dense core refers to a dense gas structure of $\sim0.1$~pc of size and $\gtrsim20$~\mo\ of mass which does not necessarily collapse into one star, but can fragment in compact condensations and form a small cluster of stars.

\begin{table}
\caption{Parameters of PdBI 1.3~mm continuum observations}
\begin{center}
{\small
\begin{tabular}{lccccc}
\noalign{\smallskip}
\hline\noalign{\smallskip}
&\multicolumn{2}{c}{Phase center}
&synth. beam
&P.A.
\\
\cline{2-3}
Source
&$\alpha (\rm J2000)$
&$\delta (\rm J2000)$
&$(''\times'')$
&($^\circ$)
&rms\,$^\mathrm{a}$\\
\noalign{\smallskip}
\hline\noalign{\smallskip}
I22172N	&22:19:08.60  	&+56:05:02.0 	&$0.47\times0.38$	&26	&0.48\\
I22134	&22:15:09.23  	&+58:49:08.9	&$0.56\times0.49$	&41	&0.33\\ 
I22198	&22:21:26.78 	&+63:51:37.6	&$0.43\times0.39$	&27	&2.0\\
A5142     	&05:30:48.02  	&+33:47:54.5  	&$0.52\times0.36$	&12	&2.8\\
\hline
\end{tabular}
\begin{list}{}{}
\item[$^\mathrm{a}$] Rms noise in m\jpb.
\end{list}
}
\end{center}
\label{tobs}
\end{table}

In Section~2 we describe the PdBI observations in the most extended (A) configuration\footnote{The PdBI A-configuration dataset of 1.3~mm continuum and CO\,(2--1) emission presented in this work allowed us to simultaneously detect emission of complex organic molecules in I22198 and A5142. This is reported in Palau \et\ (2011).}; in Section~3, we present the results of the 1.3~mm continuum and CO\,(2--1) emission for the four regions; in Section~4, we present the total sample of 18 massive dense cores (14 from literature + 4 from this work); in Section~5 we discuss the role of the global properties of massive dense cores in the fragmentation process, and compare our observations to radiation-magneto-hydrodynamic simulations of collapsing massive dense cores.
Finally, in Section~6 we summarize our main conclusions.

\begin{figure*}
\begin{center}
\begin{tabular}[b]{c}
    \epsfig{file=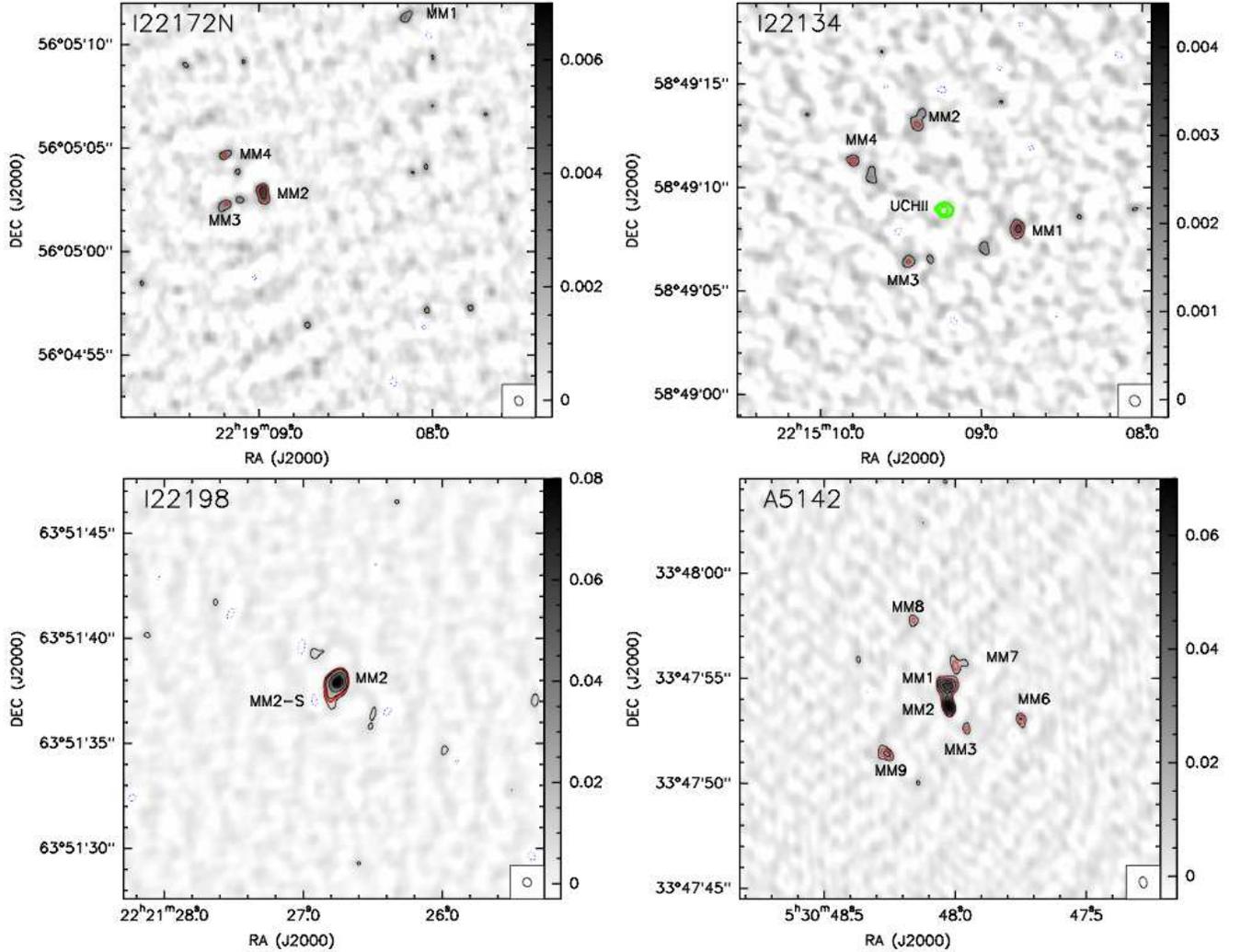, width=18cm, angle=0}\\
\end{tabular}
\caption{PdBI maps at 1.3~mm in a field of view of $20''$ ($\sim$primary beam) of diameter. 
{\bf Top-left:} I22172N. Contours are $-4$, 4, and 8  times the rms noise of the map, 0.55~m\jpb.  
Synthesized beam, shown in the bottom right corner, is  $0.47''\times0.38''$, PA$\sim26.0^\circ$.
{\bf Top-right:} I22134. Contours are $-4$, 4, and 8 times the rms noise of the map, 0.33~m\jpb. Synthesized beam is  $0.56''\times0.49''$, PA$\sim40.8^\circ$. Green contours correspond to the 3.6~cm emission tracing an UCHII region, and are $-4$, 4, 8, and 16 times 30~$\mu$\jpb\ (S\'anchez-Monge 2011).
{\bf Bottom-left:} I22198. 
Contours are  $-4$, 4, 8, 16, and 32 times the rms noise of the map, 2.0~m\jpb. 
Synthesized beam is  $0.43''\times0.39''$, PA$\sim27.1^\circ$.
{\bf Bottom-right:} A5142. Contours are $-4$, 4, 8, 12, 16, 20, and 24 times the rms noise of the map, 2.8~m\jpb. Noise is limited by dynamic range and phase noise. The synthesized beam is  $0.52''\times0.36''$, PA$\sim12.0^\circ$.
In all panels the red contour corresponds to the identification level of 6$\sigma$.
}
\label{fcont}
\end{center}
\end{figure*}

\begin{table*}
\caption{Parameters of the sources detected with the PdBI at 1.3~mm (continuum) for the 4 regions studied in this work}
\begin{center}
{\small
\begin{tabular}{lccccccccc}
\noalign{\smallskip}
\hline\noalign{\smallskip}
&\multicolumn{2}{c}{Position$^\mathrm{a}$}
&Deconv.ang.size
&P.A.
&Deconv.size
&$I_\mathrm{\nu}^\mathrm{peak}$~$^\mathrm{a}$
&$S_\mathrm{\nu}$~$^\mathrm{a}$
&Mass$^\mathrm{b}$
\\
\cline{2-3}
Source
&$\alpha (\rm J2000)$
&$\delta (\rm J2000)$
&$(''\times'')$
&($^\circ$)
&(AU$\times$AU)
&(m\jpb)
&(mJy)
&(\mo)
&Association$^\mathrm{c}$\\
\noalign{\smallskip}
\hline\noalign{\smallskip}
I22172N-MM1         	&22:19:08.155   &56:05:11.41 	&$-$                      	&$-$		&$<1000$               	&$9.0\pm0.9$	&$9\pm4$		&0.7--3	&IR\\
I22172N-MM2       	&22:19:08.974   &56:05:02.97 	&$0.94\times 0.40$	&$2$   	&$2260 \times 960$	&$6.1\pm0.5$ 	&$29\pm9$    	&2--10	&Outf\\
I22172N-MM3   	&22:19:09.178   &56:05:02.37	&$-$                      	&$-$     	&$<1000$               	&$8.1\pm0.6$ 	&$8\pm2$      	&0.6--3	&-\\
I22172N-MM4         	&22:19:09.199   &56:05:04.66 	&$-$                       	&$-$     	&$<1000$               	&$6.3\pm0.6$ 	&$6\pm2$     	&0.5--2	&-\\
\hline
I22134-MM1         	&22:15:08.778   &58:49:07.90 	&$-$                       	&$-$     	&$<1300$             	&$2.8\pm0.4$ 	&$2.8\pm1.0$	&0.2--1	&-\\ 
I22134-MM2         	&22:15:09.395   &58:49:13.04	&$-$                       	&$-$     	&$<1300$              	&$2.1\pm0.4$  	&$2.1\pm0.8$	&0.2--0.8	&-\\ 
I22134-MM3         	&22:15:09.455   &58:49:06.45 	&$-$                        	&$-$     	&$<1300$              	&$2.2\pm0.4$ 	&$2.2\pm0.7$	&0.2--0.9	&-\\ 
I22134-MM4         	&22:15:09.804   &58:49:11.33	&$-$                       	&$-$     	&$<1300$              	&$2.1\pm0.5$ 	&$2.1\pm0.8$	&0.2--0.8	&-\\ 
\hline
I22198-MM2         	&22:21:26.753   &63:51:37.90	&$0.68\times 0.39$	&$-35$	&$520 \times 300$ 	&$91.9\pm2.0$	&$250\pm70$	&0.5--2	&HC,Outf\\
I22198-MM2-S      	&22:21:26.807   &63:51:37.14	&$-$                       	&$-$     	&$<300$                	&$18.8\pm0.8$	&$18\pm5$  	&0.1--0.6	&IR,Outf?\\
\hline
A5142-MM1          	&05:30:48.031  &33:47:54.64 	&$0.68\times0.51$ 	&$-86$ 	&$1200\times900$ 	&$38.3\pm2.8$	&$210\pm60$	&2--9	&HC,Outf\\
A5142-MM2         	&05:30:48.024  &33:47:53.67 	&$0.55\times0.25$	&$18$  	&$1000\times400$	&$62.6\pm2.8$	&$150\pm40$  	&2--6	&HC,Outf\\
A5142-MM3         	&05:30:47.953  &33:47:52.57 	&$-$                       	&$-$     	&$<700$                 	&$22.5\pm1.3$	&$23\pm7$    	&1--4	&OA,W\\
A5142-MM6          	&05:30:47.749  &33:47:53.03 	&$-$                       	&$-$     	&$<700$                 	&$21.9\pm1.3$	&$22\pm6$    	&0.9--4	&OA,W\\
A5142-MM7     		&05:30:47.980  &33:47:55.53 	&$-$                       	&$-$     	&$<700$                	&$18.6\pm1.3$	&$19\pm6$    	&0.8--4	&IR?\\
A5142-MM8         	&05:30:48.152  &33:47:57.78 	&$-$                        	&$-$     	&$<700$                 	&$18.6\pm1.3$	&$19\pm5$  	&0.8--4	&-\\
A5142-MM9        	&05:30:48.257  &33:47:51.52 	&$-$                        	&$-$     	&$<700$                 	&$19.4\pm1.3$	&$20\pm6$      	&0.9--4	&OA\\
\hline
\end{tabular}
\begin{list}{}{}
\item[$^\mathrm{a}$] Position and flux density derived by fitting a point source or a Gaussian in the $uv$-plane. Peak intensity measured directly in the cleaned image. Uncertainty in the peak intensity, $\sigma$, is the rms noise of the cleaned image for the sources fitted with a Gaussian, or the rms given by the fit of a point source. 
Error in flux density has been calculated as $\sqrt{(\sigma\,\theta_\mathrm{source}/\theta_\mathrm{beam})^2+(\sigma_\mathrm{flux-scale})^2}$ (Beltr\'an \et\ 2001), where $\theta_\mathrm{source}$ and $\theta_\mathrm{beam}$ are the size of the source and the beam respectively, and $\sigma_\mathrm{flux-scale}$ is the error in the flux scale, which takes into account the uncertainty on the calibration applied to the flux density of the source ($S_\nu\times\%_\mathrm{uncertainty}$). Peak intensities and flux densities (with uncertainties) are corrected for the primary beam response.
\item[$^\mathrm{b}$] Masses derived assuming a dust temperature of 10--30~K (except for I22198-MM2, A5142-MM1 and A5142-MM2, for which we used a range of 30--100~K, Palau \et\ 2011), and a dust (+gas) mass opacity coefficient at 1.3~mm of 0.00899~cm$^2$\,g$^{-1}$ (Ossenkopf \& Henning 1994, see main text). The uncertainty in the masses due to the opacity law is estimated to be a factor of 2.
\item[$^\mathrm{c}$] Association with signposts of stellar activity: IR = infrared source; Outf = source driving outflow; OA = source along outflow axis; HC = hot molecular core (Palau \et\ 2011); W = water maser emission at velocities about 5~\kms\ away from systemic velocities (Hunter \et\ 1995; 1999); see Sect.~\ref{srco} for further details.
\end{list}
}
\end{center}
\label{tcont}
\end{table*}

%%%%%%%%%%%%%%%%%%%%%%%%%%%%%%%%%%%%%%%%%%%%%%%%%%%%%%%%%%%%%%%%%%%%%%%%%%%%%%%%
\section{Observations \label{sobs}}

The PdBI observations, partially described in Palau \et\ (2011), were carried out on 2010 January 10th with the array
in A configuration at 230.538~GHz to observe A5142. The baselines range from 136 to 760~m. The system temperature was 150--200~K and the averaged atmospheric precipitable water vapor was $\sim 2$~mm. Radio frequency bandpass calibration was carried out using B0851+202. 
Gain phases and amplitudes were calibrated by observing the nearby quasars B0552+398 and J0512+294, yielding a phase rms of 20--$60^\circ$. The adopted flux density for the flux calibrator, 3C273, was 16.57~Jy.  A subsequent track was carried out on 2010 January 18th to observe in track-sharing mode the I22172N, I22134, and I22198 regions. The system temperature was around 150~K and the precipitable water vapor was $\sim\,1$~mm. 3C454.3 was used for the radio frequency bandpass calibration. Phases were calibrated with B2146+608, and B2037+511, and we obtained a phase rms of 20--$40^\circ$. A flux of 0.23~Jy was fixed for the flux calibrator, B2146+608.  
The absolute flux density scale was determined from 3C273 and B2146+608, with an estimated uncertainty around 30\%.
% Absolute position accuracy estimate:
% From S/N: for strong sources S/N > 15 -> 0.02''
% From rms noise in phases: for 40¼ -> 0.055''
The absolute position accuracy was estimated to be around $\sim0.05''$.
A correlator unit of 40~MHz of bandwidth with 256 spectral channels was used to observe the CO\,(2--1) line in each polarization.
Three additional units (per polarization) of 320~MHz bandwidth with 128 channels were used to observe the continuum across $\sim1$~GHz.

Calibration and imaging were performed using the \textsc{Clic} and \textsc{Mapping}\footnote{http://www.iram.fr/IRAMFR/GILDAS} packages of \textsc{Gildas} software, using the standard procedures. The continuum $uv$-data of I22134 and the CO(2--1) $uv$-data of all the four sources were tapered to 400~m baselines ($uv$-distance where the weight of the taper Gaussian is 37\%), to increase the signal-to-noise (S/N) ratio.
For the CO\,(2--1), we smoothed the channel maps to a velocity resolution of 0.6~\kms\ per channel.
Typical rms noises (in line-free channels) per channel are 20--30~m\jpb. 
The final continuum maps were built by carefully selecting the channels of the 320~MHz units which were free of line emission.
The phase centers, synthesized beam sizes, and rms of the final cleaned continuum maps 
are listed in Table~\ref{tobs}. 
I22198 and A5142 images are limited by dynamic range and A5142 additionally has phase noise.
The primary beam of the PdBI at  1.3~mm is $22''$, allowing us to cover a field of view of 0.1--0.2~pc.  

%coordinates and Vlsr:
%afgl5142  eq 2000 05:30:48.02 33:47:54.5 LSR -3.9
%i22172    eq 2000 22:19:08.6  56:05:02.0 LSR -43.8	
%i22134	   eq 2000 22:15:09.23 58:49:08.9 LSR -18.3
%i22198    eq 2000 22:21:26.78 63:51:37.6 LSR -10.5 

%%%%%%%%%%%%%%%%%%%%%%%%%%%%%%%%%%%%%%%%%%%%%%%%%%%%%%%%%%%%%%%%%%%%%%%%%%%%%%%%
\section{Results \label{sres}}

\subsection{1.3 mm continuum \label{srcont}}

In Fig.~\ref{fcont} we present the cleaned images of the four regions observed with the PdBI in A configuration (hereafter, PdBI-A). Sources above $6\sigma$ (with $\sigma$ the rms noise level) were identified in the final cleaned images, and are listed in Table~\ref{tcont}. In order to confirm that the $6\sigma$ identifications in the cleaned images are well detected also in the $uv$-plane, we fitted in the $uv$-plane a Gaussian or a point-like source to the strongest source identified in the cleaned map, and subtract the fitted source to the observed visibilities ($uv$-plane). Then, we fitted in the residual visibilities the second strongest source identified in the cleaned map, etc. All the identified sources (in the cleaned map) were detected in the $uv$-plane with a S/N ratio larger than 10 (except for I22134, for which S/N in the $uv$-plane was $>4.6$), giving us confidence on the identification method.
Position, flux densities and sizes for each source were determined from the fits of a Gaussian or a point-like source in the $uv$-plane. These parameters are listed in Table~\ref{tcont}, together with an estimate of the mass for each millimeter source.

In the I22172N field we detected four sources: three faint point sources and a strong extended source, MM2, which is elongated roughly in the north-south direction, with a size of $2300\times1000$~AU, and a mass of 2--10~\mo\ (Table~\ref{tcont}). The mass is derived assuming a dust temperature of 10--30~K, and a dust mass opacity coefficient at 1.3~mm of 0.899 cm$^2$\,g$^{-1}$ (column 6 of Table~1 of Ossenkopf \& Henning 1994, corresponding to agglomerated dust grains with thin ice mantles at densities $\sim10^6$~\cmt). We estimated an uncertainty in the masses due to uncertainty in the dust opacity of about a factor of 2.
MM3 and MM4 fall within $3''$ to the (north)east ($<7200$~AU) of MM2, and are coinciding with a \nh\ emission peak (for which a rotational temperature of $\sim20$~K is estimated, S\'anchez-Monge \et, in prep.), while MM1 is $\sim12''$ to the northwest of MM2. For MM1, MM3 and MM4, we estimated masses around 0.5--3~\mo, making the same assumptions given above.

Towards I22134 we detected four point sources with a flux of $\sim2$~mJy, located in all directions and at similar distances, $4''$ or $\sim10000$~AU, from the UCHII region. The masses of the fragments are $\sim0.7$~\mo, and the UCHII region is not detected at 1~mm. Such a non-detection is consistent with the extrapolation of the flux density of the centimeter source at 1.3~cm ($\sim0.8$~m\jpb) to 1.3~mm, using the measured spectral index of $-0.18$ (S\'anchez-Monge 2011). 

For I22198 we detected one strong source, MM2 (following the labeling in S\'anchez-Monge \et\ 2010), extended in the southeast-northwest direction, with a faint extension at 6$\sigma$, MM2-S, $0.8''$ (600~AU) to the southeast, already reported in Palau \et\ (2011). MM2 has a size of $500\times300$~AU at a PA=$-35$\degr, and its flux density, of $\sim250$~mJy, corresponds to a mass of $\sim0.5$--2~\mo. Here we assumed a dust temperature range of 30--100~K because MM2 is associated with emission from complex organic molecules tracing a hot core at $\sim100$~K (Palau \et\ 2011). The faint extension MM2-S has a flux density corresponding to a mass of $\sim0.5$~\mo\ (assuming a dust temperature of 10--30~K).

Finally, for the case of A5142, the millimeter emission is dominated by two partially extended and strong sources, MM1 and MM2, separated by $\sim1''$ or 1800~AU, which are surrounded by five faint point-like sources (labeled following Zhang \et\ 2007). The masses of MM1 and MM2 are 2--9~\mo\ (assuming, as in the case of I22198-MM2, a dust temperature of 30--100~K,
Palau \et\ 2011), and both sources are associated with water masers. MM1 is elongated in the east-west direction, with a size of  $1200\times900$~AU, while MM2 is oriented in the north-south direction and is sligthly smaller, of $1000\times400$~AU.
Concerning the other sources, they are compact and faint, with masses estimated around 1--4~\mo; MM7 lies at about $1''$ to the northwest of MM1, and MM3 and MM6 are associated with water maser emission (Hunter et al. 1995, 1999).

\subsection{CO\,(2--1) emission, 2MASS and Spitzer/IRAC sources \label{srco}}

\begin{figure}
\begin{center}
\begin{tabular}[b]{cc}
    \epsfig{file=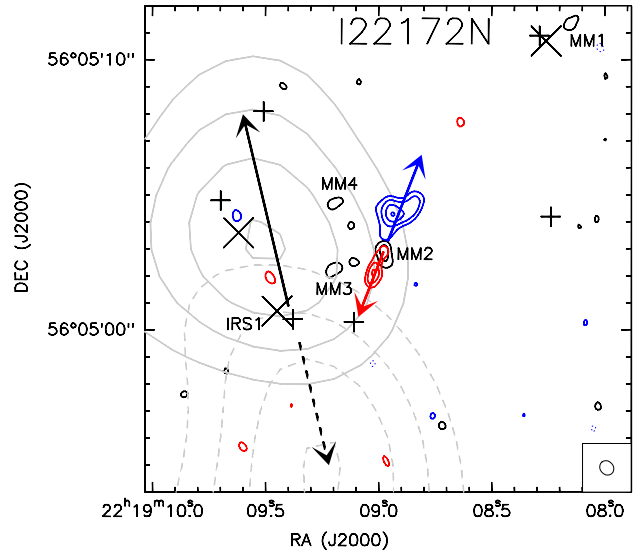, width=8.2cm, angle=0}\\
\end{tabular}
\caption{I22172N: PdBI-A map of the continuum at 1.3~mm (black contours as in Fig.~\ref{fcont}) with the CO\,(2--1) high-velocity blueshifted (blue contours, integrated from $-52.4$ to $-44.6$~\kms) and redshifted (red contours, integrated from $-40.4$ to $-36.8$~\kms) emission overplotted (systemic velocity is $-43.8$~\kms). Red and blue arrows mark the directions of the outflows seen with the PdBI-A. Grey (dashed, solid) contours and black arrows indicate the CO\,(1--0) outflow previously reported by Fontani \et\ (2004, with a beam $\sim 6''$). Contours of the PdBI CO\,(2--1) emission start at 55\% and increase in steps of 15\% of the maximum, 2.13~\jpb\,\kms, for the blueshifted emission, and start at 60\% increasing in steps of 20\% of the maximum, 0.87~\jpb\,\kms, for the redshifted emission.
Plus signs correspond to 2MASS sources, and crosses to Spitzer/IRAC sources at 3.6~\mum. The synthesized beam of the PdBI CO\,(2--1) emission, of $0.57''\times0.49''$ with P.A.= 39.2\degr, is shown in the bottom-right corner. 
}
\label{fi22172co}
\end{center}
\end{figure}

\begin{figure}
\begin{center}
\begin{tabular}[b]{cc}
    \epsfig{file=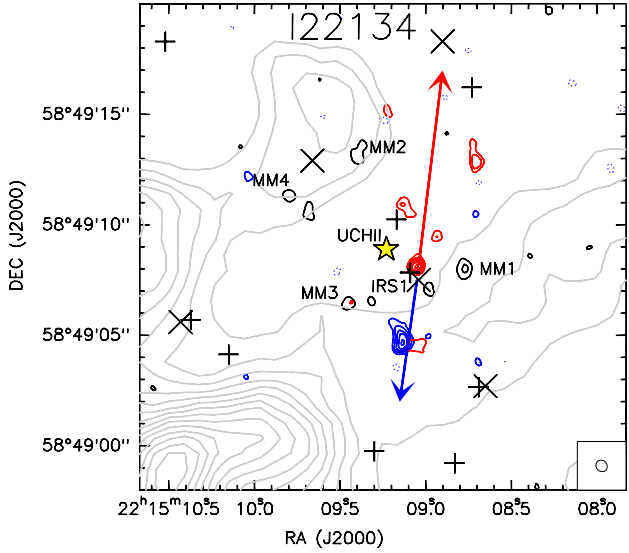, width=8.2cm, angle=0}\\
\end{tabular}
\caption{Idem as Fig.~\ref{fi22172co} for I22134. Grey contours correspond to the integrated emission of \nh(1,1) (Busquet 2010, S\'anchez-Monge 2011). Contours of the PdBI CO\,(2--1) emission start at 45\% and increase in steps of 13\% of the maximum, 1.98 and 2.16~\jpb\,\kms, for the blueshifted and redshifted emission, respectively. CO\,(2--1) blueshifted emission has been integrated from $-31.7$ to $-22.1$~\kms, while CO\,(2--1) redshifted emission has been integrated from $-14.9$ to $-4.7$~\kms\ (systemic velocity is $-18.3$~\kms). The star indicates the position of the UCHII region (Fig.~\ref{fcont} and S\'anchez-Monge 2011). The synthesized beam of the PdBI CO\,(2--1) emission, of $0.55''\times0.49''$ with P.A.= 42.5\degr, is shown in the bottom-right corner. 
}
\label{fi22134co}
\end{center}
\end{figure}

\begin{figure}
\begin{center}
\begin{tabular}[b]{c}
    \epsfig{file=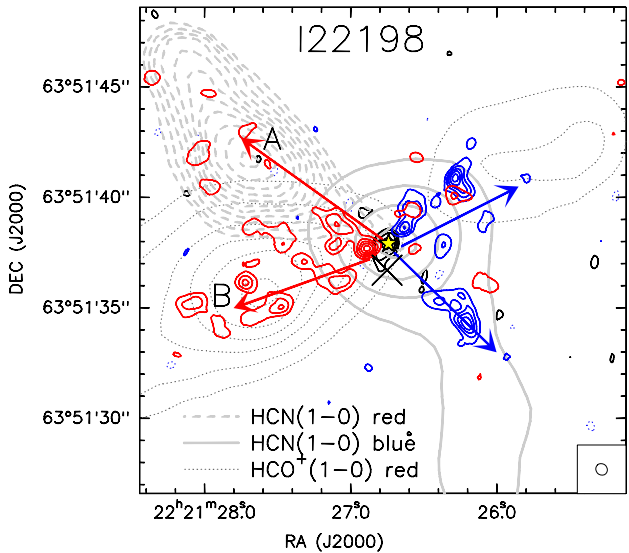, width=8.2cm, angle=0}\\    
\end{tabular}
\caption{Same as Fig.~\ref{fi22172co} for I22198. Grey (dashed, solid) contours correspond to (red-, blue-shifted) HCN\,(1--0) outflow emission; and the grey dotted contours correspond to the redshifted HCO$^+$\,(1--0) outflow emission (beam around $3''$, S\'anchez-Monge 2011).  Contours of the PdBI CO\,(2--1) emission start at 18\% and increase in steps of 18\% of the maximum, 1.75(1.77)~\jpb\,\kms, for the blueshifted(redshifted) emission. CO\,(2--1) blueshifted emission has been integrated from $-31.0$ to $-19.7$~\kms, while CO\,(2--1) redshifted emission has been integrated from $-5.3$ to $7.9$~\kms\ (systemic velocity is $-10.5$~\kms).
The star marks the position of a hot molecular core (S\'anchez-Monge \et\ 2010; Palau \et\ 2011), 
and the cross indicates the position of the 3.6~\mum\ Spitzer/IRAC source found in the field. The synthesized beam of the PdBI CO\,(2--1) emission, of $0.55''\times0.50''$ with P.A.= 42.5\degr,  is shown in the bottom-right corner. 
}
\label{fi22198co}
\end{center}
\end{figure}

\begin{figure*}
\begin{center}
\begin{tabular}[b]{cc}
    \epsfig{file=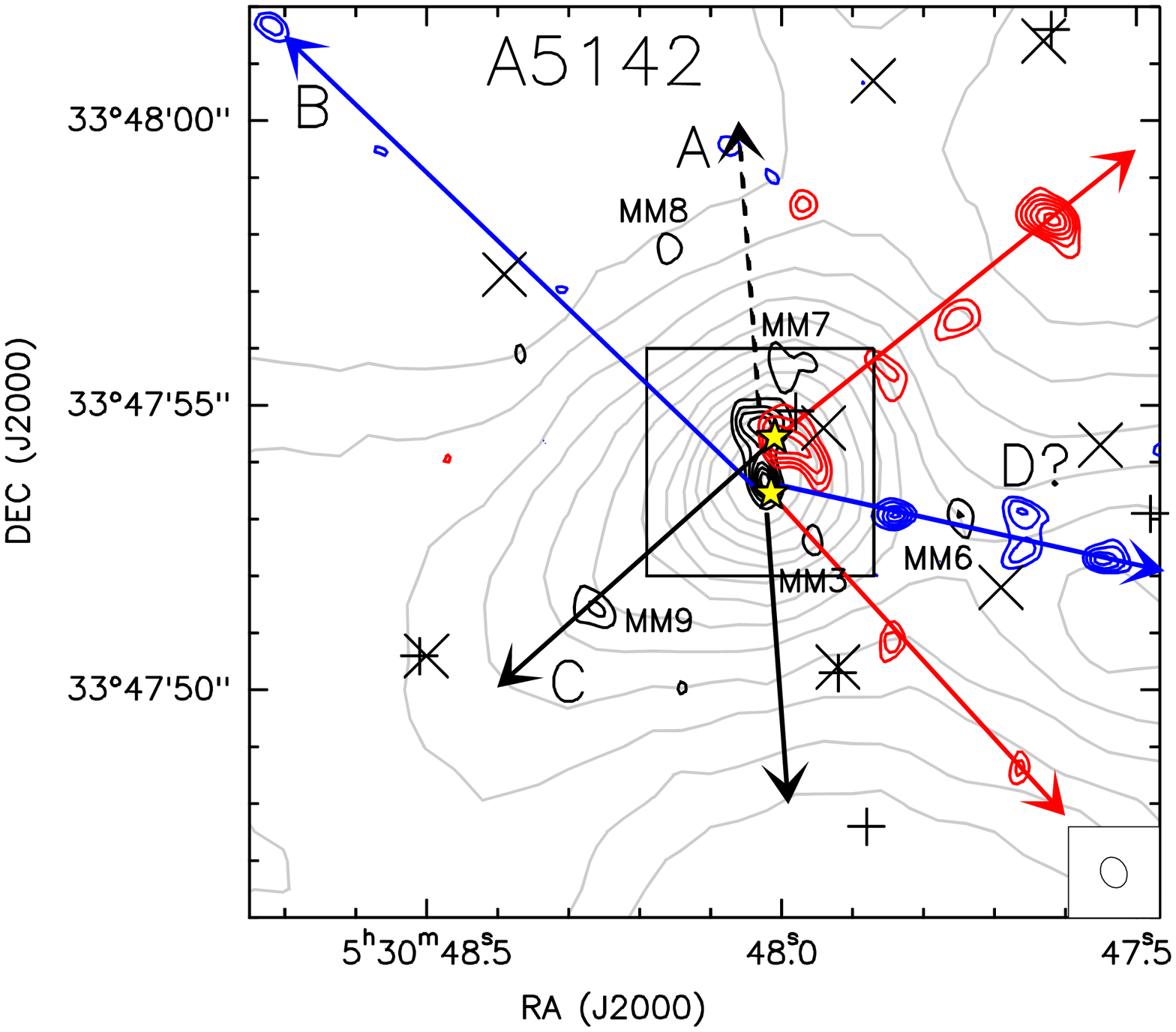, width=8.2cm, angle=0}&
    \epsfig{file=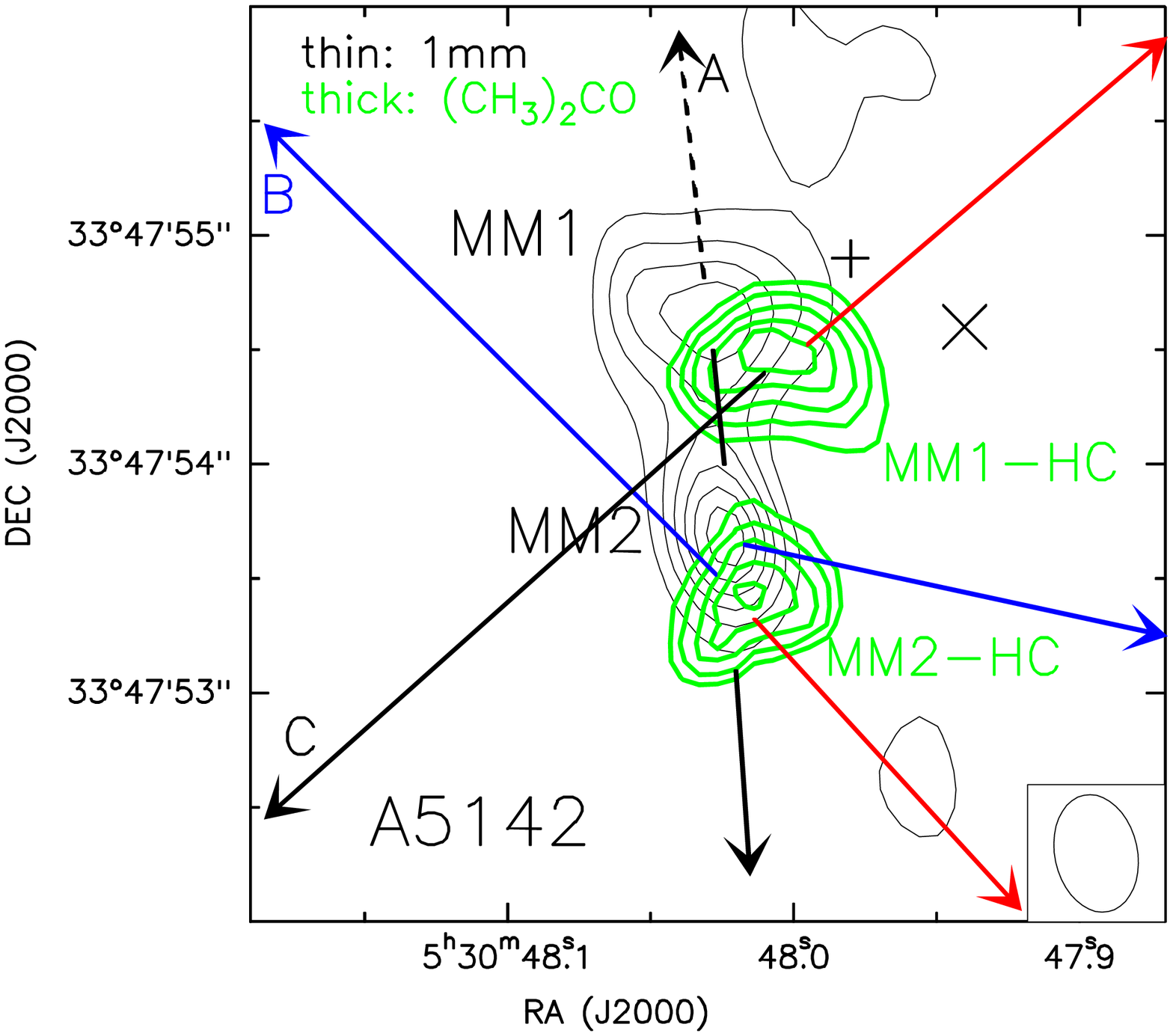, width=8.2cm, angle=0}\\
\end{tabular}
\caption{{\bf Left:} Same as Fig.~\ref{fi22134co} for A5142 (grey contours correspond to the integrated emission of \nh(1,1), Zhang \et\ 2002). 
Contours of the PdBI CO\,(2--1) emission start at 55\% and increase in steps of 10\% of the maximum, 1.85~\jpb\,\kms, for the blueshifted emission, and start at 45\% increasing in steps of 10\% of the maximum, 3.18~\jpb\,\kms, for the redshifted emission. 
CO\,(2--1) blueshifted emission has been integrated from $-19.7$ to $-8.9$~\kms, while CO\,(2--1) redshifted emission has been integrated from $0.7$ to $12.7$~\kms\ (systemic velocity is $-3.9$~\kms).
Stars mark the positions of the hot molecular cores (Palau \et\ 2011), shown on the right panel; plus signs indicate the positions of 2MASS sources and crosses indicate the positions of the 3.6~\mum\ Spitzer/IRAC sources. 
{\bf Right:} zoom on the central $4''$ of A5142. Thin black contours correspond to the 1.3~mm emission shown on the left panel. Thick green contours correspond to emission of (CH$_3$)$_2$CO tracing the hot molecular cores (Palau \et\ 2011), labeled MM1-HC (northern) and MM2-HC (southern).
The synthesized beam of the PdBI CO\,(2--1) emission, of $0.56''\times0.45''$ with P.A.= 24.9\degr,  is shown in the bottom-right corner.  
}
\label{fa5142co}
\end{center}
\end{figure*}

In Figs.~\ref{fi22172co}, \ref{fi22134co}, \ref{fi22198co}, and \ref{fa5142co} we present the 1.3~mm continuum emission as in Fig.~\ref{fcont} with the high-velocity blueshifted and redshifted CO\,(2--1) emission superimposed. 
The CO\,(2--1) emission is heavily filtered out (the PdBI operating at 1~mm in A configuration is not sensitive to structures larger than $\gtrsim1''$), and reveals only the high-velocity most compact knots.
As an example, we estimated that the PdBI-A misses about 60--70\% of the high velocity CO\,(2--1) emission in I22198 (see below) compared to the flux measured with the SMA in compact configuration (S\'anchez-Monge \et\ 2010).
Although suffering from strong filtering, most of the high-velocity knots recovered with the PdBI-A are consistent with the previously known outflows in each region, and for I22172N the CO\,(2--1) emission reveals one new outflow.

In order to further understand the nature of the 1~mm continuum sources, we searched for 2MASS and Spitzer/IRAC sources (at 3.6~\mum) within the $22''$ PdBI field of view, and the positions are marked in Figs.~\ref{fi22172co}, \ref{fi22134co}, \ref{fi22198co}, and \ref{fa5142co}. The positions of the Spitzer/IRAC sources were extracted from the image at 3.6~\mum\ using the \textsc{Source-Extractor} software (Bertin \& Arnouts 1996).
In the last column of Table~\ref{tcont} we list which millimeter sources are associated with infrared emission, water maser emission (Hunter \et\ 1995, 1999), or hot molecular cores (Palau \et\ 2011). We also indicate if the sources are driving outflows or are located along outflow axes. Those millimeter sources with no evidences of stellar activity could either be starless or protostellar. We favor the protostellar option because starless cores have flat column density profiles with typical (half-power) radius of about 3000--6000~AU (\eg\ Motte \et\ 1998; Kandori \et\ 2005), implying that the PdBI-A should recover a fraction of correlated flux for these cores of only $<0.1$\% (following Palau \et\ 2010). Thus, starless cores are presumably filtered out in our observations.
In the following, we give a small description of the CO\,(2--1) PdBI-A emission for each source, together with a brief interpretation of the nature of each millimeter source.
 
\paragraph{I22172N:} The first interferometric observations studying the 3~mm and CO\,(1--0) emission in I22172N were carried out with OVRO by Fontani \et\ (2004), who find one millimeter condensation with one outflow displaced about $4''$ to the east. Our PdBI-A observations reveal one strong and elongated source, MM2, coincident with the peak of emission seen at 3~mm by Fontani \et\ (2004), and which seems to be driving a small outflow not reported before (Fig.~\ref{fi22172co}). The blueshifted lobe of this small outflow is associated with a faint H$_2$ feature reported in figure~15-bottom of Fontani \et\ (2004), which is elongated also in the southeast-northwest direction, confirming the presence of this outflow. MM2 and its nearby sources MM3 and MM4 are deeply embedded in a dense core traced in \nh(1,1) (S\'anchez-Monge et al., in prep.) and lie in a region with no near-infrared emission (see figures 10 and 15 of Fontani \et\ 2004). On the other hand, the large-scale CO\,(1--0) outflow found by Fontani \et\ (2004) seems to be driven by a near-infrared and Spitzer/IRAC source, labeled IRS1 in the figure, which falls exactly at the center of symmetry of the two CO\,(1--0) outflow lobes but which has no dust associated.  
Concerning MM1, it falls $1''$ to the northwest of one bright near-infrared and Spitzer/IRAC source. Since the positional uncertainty of Spitzer/IRAC sources is also $\sim1''$, the infrared source and MM1 are possibly associated. The faint detection of MM1 with the PdBI-A, together with its strong intensity in the near-infrared, and the lack of a strong outflow suggests that MM1 is a slightly more evolved object than IRS1 and MM2 (MM3 and MM4), as expected if star formation in I22172N is triggered by the WR star HD\,211853, which is about $5'$ to the (north)west (Liu \et\ 2012).
Overall, I22172N harbors a rich variety of young stellar objects: two protostellar candidates detected only in the millimeter range (MM3 and MM4), one confirmed protostellar object with no infrared emission and driving a small outflow (MM2), one protostellar object detected only in the infrared and driving a larger outflow (IRS1), and one protostellar object detected in the millimeter and infrared, with no outflow (MM1).
%NOTE: for I22172N we checked the VLA cm emission at 3.6 cm and there is a source detected at the position of the bright rim..., with a beam of $10''$.

\paragraph{I22134:} The \nh(1,1) and (2,2) emission of this region is studied in Busquet (2010) and S\'anchez-Monge (2011), who find that the dense gas is mainly concentrated towards the southeast of the UCHII region, with faint \nh\ emission forming a tilted `U' structure around the UCHII region (Fig.~\ref{fi22134co}), where the broadest linewidths and highest rotational temperatures are found. Thus, the dense gas seems to be tracing the walls of a cavity which could have been created by the UCHII region, similar to the cavity recently found by Fontani \et\ (2012) in IRAS\,20343+4129.  Interestingly, the millimeter sources found with the PdBI-A are distributed along the `U' \nh\ structure, at the edges facing the UCHII region, suggesting that they could be dust condensations affected by the expanding ionization front creating the cavity.   
As for the CO\,(2--1) emission, all the previous observations are done with single-dish and mainly trace the large-scale CO structure (\eg\ Beuther \et\ 2002b), which cannot be related to the knots found with the PdBI-A. We find one main redshifted knot, in between MM1 and MM3, and one strong blueshifted knot about $3''$ the south. Only $0\farcs5$ to the south of the redshifted CO knot there is a 2MASS and Spitzer/IRAC source, as well as emission at 4$\sigma$ detected with PdBI-A at 1~mm. All this suggests that this infrared source, labeled IRS1 in Fig.~\ref{fi22134co}, could be the driving source of an outflow elongated in the north-south direction. Observations in the CD configuration, recovering larger angular structures, are required to confirm the existence of such an outflow. Thus, in this region we found 4 millimeter sources with no infrared emission plus one faint millimeter source being bright in the infrared, with the UCHII region located at the center of all these 5 young stellar objects.

\begin{table*}
\caption{Massive dense cores studied with interferometers at 1.3mm with high sensitivity and down to a spatial resolution $\lesssim1000$~AU}
\begin{center}
{\small
\begin{tabular}{lcrrccccccccc}
\noalign{\smallskip}
\hline\noalign{\smallskip}
&D
&$L_\mathrm{bol}$
&$T_\mathrm{bol}$\supb\
&Rms\supc
&$M_\mathrm{min}$\supc\ 
&Spat. res.\supd
&LAS\supd\ 
&&&
&st. dens.\supf
&separation\supf
\\
ID-Source\supa
&(kpc)
&(\lo)
&(K)
&(mJy)
&(\mo)
&(AU)
&(AU)
&$N_\mathrm{mm}$\supe\ 
&$N_\mathrm{mm+IR}$\supe\ 
&$N_\mathrm{IR}/N_\mathrm{mm}$
&($10^4$\,pc$^{-3}$)
&(AU)
\\
\noalign{\smallskip}
\hline\noalign{\smallskip}
1-IC1396N       		&0.75	&290	&52		&0.52	&0.03	&340    	&900	&4	&8		&1.00	&69	&1800\\
2-I22198        		&0.76	&340	&59		&2.0		&0.13	&300	&700	&1.5	&2		&0.33	&3	&$-$\\
3-NGC2071-IRS1\supg\ &0.42	&440	&98		&0.50	&0.20	&200	&5500	&4	&14		&2.50	&16	&2900\\%use comp config.
4-NGC7129-FIRS2	&1.25	&460       	&58		&2.9		&0.49	&750     	&1500	&1	&2		&1.00	&3	&$-$\\
5-CB3-mm		&2.50	&700	&49		&0.43	&0.29	&875	&2500	&2	&$\gtrsim2$&$\gtrsim0.00$&$\gtrsim3$&$-$\\
6-I22172N-IRS1      	&2.40	&830     	&195	&0.55	&0.34	&1000    	&2100	&3	&6		&1.00	&6	&4400\\
7-OMC-1S-136        	&0.45	&2000  	&$-$		&30		&0.66	&540 	&3300	&9	&21		&1.33	&9	&3000\\
8-A5142              	&1.80	&2200     	&55		&2.8		&0.99	&700  	&1600	&7	&11		&0.57	&7	&3700\\
9-I05358+3543NE	&1.80	&3100   	&67		&1.5		&0.53	&900     	&2100	&4	&7		&0.75	&10	&3500\\
10-I20126+4104	&1.64	&8900	&61		&2.6		&0.76	&1400     	&4000	&1	&3		&2.00	&23	&2700\\
11-I22134-IRS1      	&2.60	&11800  	&93		&0.30	&0.22	&1300      	&2300	&3.5	&7 	 	&1.00	&4	&4600\\ 
12-HH80-81         	&1.70	&21900	&81		&3.0		&0.94	&830       	&4000	&3	&6 		&1.00	&6	&4200\\
13-W3IRS5          	&1.95	&140000	&114	&1.2		&0.50	&720  	&2300	&3.5	&$\gtrsim$5.5	&$\gtrsim$0.57	&$\gtrsim46$&2300\\
14-AFGL\,2591		&3.00	&190000	&226	&0.51	&0.49	&1000	&1900	&1.5	&$\gtrsim$1.5	&$\gtrsim$0.00	&$\gtrsim3$&$-$\\
\hline
%16-CygX-N3		&1.40	&100	&103	&1.2		&0.26	&1400	&6900	&3	&6		&1.00	&	&\\
15-CygX-N53		&1.40	&300	&33		&1.9		&0.41	&1400	&6900	&4	&7		&0.75	&11	&3400\\
16-CygX-N12		&1.40	&320	&58		&1.9		&0.41	&1400	&6900	&2.5	&5		&1.00	&23	&2800\\
17-CygX-N63		&1.40	&470	&39		&4.2		&0.90	&1400	&6900	&2.5	&4		&0.60	&72	&2000\\
18-CygX-N48		&1.40	&4400	&48		&2.2		&0.47	&1400	&6900	&4	&5		&0.25	&11	&3600\\
\hline
\end{tabular}
\begin{list}{}{}
\item[$^\mathrm{a}$] `I' in front of a source name stands for `IRAS'. References: 1: Neri \et\ (2007); 2: this work; 3: Carrasco-Gonz\'alez \et\ (2012); 4: Fuente \et\ (2005); 5: Fuente \et\ (2007); 6: this work; 7: Zapata \et\ (2005); 8: this work; 9: Leurini \et\ (2007); 10: Cesaroni \et\ (2005), Moscadelli \et\ (2011); 11: this work; 12: Fern\'andez-L\'opez \et\ (2011); 13: Rod\'on \et\ (2008); 14: Wang \et\ (2012a), Jim\'enez-Serra \et\ (2012); 15--18: Bontemps \et\ (2010).
\item[$^\mathrm{b}$]  Bolometric temperature, defined as the temperature of a blackbody having the same mean frequency as the observed spectral energy distribution (\eg\ Chen \et\ 1995).
\item[$^\mathrm{c}$] Rms of the millimeter images compiled for this work (see tablenote `a' for references). $M_\mathrm{min}$ is the mass sensitivity derived assuming 5 times the rms noise, a dust temperature of 20~K, and a dust mass opacity coefficient at 1.3~mm from Ossenkopf \& Henning (1994) as in Table~\ref{tcont}. 
\item[$^\mathrm{d}$] The spatial resolution is calculated from the synthesized beam of the millimeter images compiled for this work (see tablenote `a' for references), and the adopted distance for each core.
LAS is the largest angular scale detectable by the interferometric observations, estimated following Palau \et\ (2010) and converted to spatial scale using the adopted distances. 
\item[$^\mathrm{e}$] $N_\mathrm{mm}$ is defined as the number of millimeter sources within $\sim0.1$~pc (in diameter, see Section~\ref{saNmm}). $N_\mathrm{mm+IR}=N_\mathrm{mm}+N_\mathrm{IR}$, with $N_\mathrm{IR}$ being the number of infrared sources (2MASS + Spitzer/IRAC at 3.6~\mum) without millimeter counterpart within $\sim0.1$~pc. 
In case of 4 or 5$\sigma$ detection but with independent claims in the literature of being a true source, we counted the source has `half' (0.5). Positional uncertainties are adopted as $\sim2''$ for Spitzer/IRAC, and $\sim1''$ for 2MASS and millimeter sources, and two sources were considered to be associated if their angular separation is smaller than the sum of the positional uncertainties.
\item[$^\mathrm{f}$] First approach to (proto)stellar density estimated from $N_\mathrm{mm+IR}$ divided by the volume of a sphere with the same projected area onto the sky as the polygon with minimum area ($A_\mathrm{poly}$) encompassing all the sources. For the case of $N_\mathrm{mm+IR}\leqslant2$, we used the average spatial area measured for regions with $N_\mathrm{mm+IR}>2$. The separation is an `average' projected separation between sources in each region, estimated as $\sqrt{A_\mathrm{poly}/N_\mathrm{mm+IR}}$.
\item[$^\mathrm{g}$] Observations of 3~mm continuum emission, and we assumed a spectral index of 3 to estimate the 1~mm flux densities. 
\end{list}
}
\end{center}
\label{tstatistics}
\end{table*}

\begin{table*}
\caption{Properties of massive dense cores studied with interferometers at 1.3~mm with high sensitivity and down to a spatial resolution $\lesssim1000$~AU}
\begin{center}
{\small
\begin{tabular}{lcccccccccccc}
\noalign{\smallskip}
\hline\noalign{\smallskip}
&$M_\mathrm{SD}$\supb
&$L_\mathrm{bol}/M_\mathrm{SD}$
&Size\supc
&$\Sigma$\supc
&$n_\mathrm{H_2}$\supc
&$M_\mathrm{max}$\supd
&CFE\supe
&$M_\mathrm{Jeans}^\mathrm{th}$\,\supf
&$\sigma_\mathrm{no-th}$\supg
&$M_\mathrm{Jeans}^\mathrm{tot}$\,\supf
&
\\
ID-Source\supa\
&(\mo)
&(\lo/\mo)
&(AU)
&(g\,cm$^{-2}$)
&($10^5$\,cm$^{-3}$)
&(\mo)
&(\%)
&(\mo) 
&(\kms)
&(\mo)
&$\beta_\mathrm{rot}$\suph\
\\
\noalign{\smallskip}
\hline\noalign{\smallskip}
1-IC1396N       		&36       	&8		&15700	&1.64	&2.8		&0.1    	&3 	 	&1.1		&$-$		&$-$		&0.016\\
2-I22198        		&17      	&20		&5300	&6.75	&34		&1.0		&20  	&0.3		&0.50	&7		&0.003\\
3-NGC2071-IRS1	&39		&11		&9400	&4.96	&14		&0.7		&10		&0.5		&0.41	&8		&0.066\\
4-NGC7129-FIRS2	&59       	&8		&12500	&4.27	&9.1		&1.8     	&9  		&0.6		&$-$		&$-$		&$-$\\
5-CB3-mm		&140	&5		&40300	&0.96	&0.6		&1.0		&3		&2.2		&$-$		&$-$		&$-$\\
6-I22172N-IRS1      	&150     	&6		&46500	&0.76	&0.4		&1.1    	&4  		&2.7		&0.59	&96		&0.062\\
7-OMC-1S-136      	&120  	&17		&21500	&2.89	&3.6		&1.8 	&35     	&0.9		&$-$		&$-$		&$-$	\\
8-A5142              	&280     	&8		&18100	&9.65	&14		&5.0  	&12  	&0.5		&0.63	&20		&0.011\\
9-I05358+3543NE	&380   	&8		&34900	&3.57	&2.7		&1.0     	&3     	&1.1		&0.50	&26		&0.005\\
10-I20126+4104	&200	&44		&31800	&2.25	&1.9		&4.3    	&6		&1.3		&0.37	&16		&0.004\\
11-I22134-IRS1      	&250  	&47		&65300	&0.67	&0.3		&0.1      	&0.5   	&3.4		&0.33	&34		&0.072\\ 
12-HH80-81         	&360	&60		&32900	&3.80	&3.1		&9.3      	&9      	&1.0		&0.67	&51		&0.041\\ 	
13-W3IRS5          	&2500	&56		&59100	&8.03	&3.6		&2.2  	&$\gtrsim0.4$&0.9	&$-$		&$-$		&0.021\\
14-AFGL\,2591		&1300	&140	&83300	&2.19	&0.7		&3.9		&0.9		&2.1		&$-$		&$-$		&0.013\\
\hline
15-CygX-N53\supi	&68		&4		&19700	&1.98	&2.7		&$\lesssim2.2$	&$\lesssim27$	&1.1	&$-$	&$-$		&$-$	\\
16-CygX-N12\supi	&70		&5		&16800	&2.79	&4.4		&$\lesssim1.3$	&$\lesssim16$	&0.8	&$-$	&$-$		&$-$	\\
17-CygX-N63\supi	&47		&10		&10100	&5.18	&14		&$\lesssim6.0$	&$\lesssim60$	&0.5	&$-$	&$-$		&$-$	\\
18-CygX-N48\supi	&160	&28		&22500	&3.54	&4.2		&$\lesssim1.3$	&$\lesssim8$	&0.9	&$-$	&$-$		&$-$	\\
\hline
\end{tabular}
\begin{list}{}{}
\item[$^\mathrm{a}$] `I' in front of a source name stands for `IRAS'.
\item[$^\mathrm{b}$] Mass of the dense core measured with the JCMT at 850~\mum\ or the IRAM\,30m at 1.2~mm (Jenness \et\ 1995; Motte \et\ 2007; di Francesco \et\ 2008), for a dust temperature of 20~K, and a dust (+gas) mass opacity coefficient of 0.0175\,cm$^2$\,g$^{-1}$ at 850~\mum\ and 0.00899\,cm$^2$\,g$^{-1}$ at 1.2~mm (Ossenkopf \& Henning 1994).
\item[$^\mathrm{c}$] Core size is the deconvolved FWHM. Surface density ($\Sigma$) and mean particle density ($n$) of the dense core are estimated assuming spherical symmetry and using $\Sigma=M_\mathrm{SD}/(\pi\,R^2)$ (with $R=FWHM/2$) and $n=3\,M_\mathrm{SD}/(4\pi\,FWHM)$ (Motte \et\ 2007).
\item[$^\mathrm{d}$] $M_\mathrm{max}$ is the mass of the strongest millimeter condensation measured by an interferometer with a very extended configuration (see Table~\ref{tstatistics} for the references), assuming a dust temperature of 50~K and the dust opacity law of Ossenkopf \& Henning (1994) as in Table~\ref{tcont}. The uncertainty in the mass estimate due to the temperature assumption is a factor of 2.
\item[$^\mathrm{e}$] The Core Formation Efficiency is calculated from the ratio of the sum of the masses of all the fragments measured with an interferometer in very extended configuration (Table~\ref{tstatistics}) and the single-dish mass derived from the peak intensity, making the same assumptions as tablenote `b'. Lower limits correspond to single-dish masses possibly contaminated by other nearby cores.
\item[$^\mathrm{f}$] $M_\mathrm{Jeans}^\mathrm{th}$ is the thermal Jeans mass for a temperature of 20~K
($M_\mathrm{Jeans}^\mathrm{th}=0.6285\,[T/10\mathrm{K}]^{3/2}[n_\mathrm{H_2}/10^5\mathrm{cm}^{-3}]^{-1/2}$~\mo).
$M_\mathrm{Jeans}^\mathrm{tot}$ is the Jeans mass estimated using the total velocity dispersion (thermal+non-thermal, Pillai \et\ 2011). 
\item[$^\mathrm{g}$] $\sigma_\mathrm{non-th}$ is the non-thermal velocity dispersion, calculated by fitting the hyperfine structure of \nh(1,1) emission of a quiescent core in the surroundings ($\lesssim2'$) of our studied massive dense core (see references in Section~\ref{saprop}), and assuming a thermal dispersion corresponding to 20~K. 
\item[$^\mathrm{h}$] $\beta_\mathrm{rot}$: ratio of rotational-to-gravitational energy, as defined in Chen \et\ (2012, see Sect.~\ref{saprop}), derived from dense gas tracers tracing the massive dense core at scales of $5''$--$10''$.
\item[$^\mathrm{i}$] Regions from Bontemps \et\ (2010) observed with the PdBI in CD configuration. This implies that some of the parameters are upper limits when compared to the rest of the sample (observed with more extended configurations). The parameters are essentially the same as reported in Bontemps \et\ (2010), using the opacity law adopted in this work and the new distance of Cygnus\,X, 1.4~kpc (Rygl \et\ 2012).
\end{list}
}
\end{center}
\label{tstatistics2}
\end{table*}

\paragraph{I22198:} 
The PdBI-A CO\,(2--1) emission in I22198 is resolved into several knots which are mainly aligned along the (south)east-(north)west direction, with additional faint redshifted knots to the northeast of MM2 and one strong blueshifted knot about $5''$ to the southwest of MM2 (Fig.~\ref{fi22198co}). This could be interpreted as one single outflow with a wide opening angle, or as two independent outflows, outflow `A' and `B' as marked in Fig.~\ref{fi22198co}. Observations of the high-velocity emission of different outflow tracers (carried out with the PdBI in CD configuration, S\'anchez-Monge 2011) seem to favor the two-outflow scenario. As shown in grey contours in Fig.~\ref{fi22198co}, the HCN\,(1--0) high-velocity emission is bipolar, centered on MM2, and aligned in the southwest-northeast direction. Similarly to HCN, the SiO\,(2--1) high-velocity emission is only found elongated to the southwest of MM2 (see figure~1 of S\'anchez-Monge \et\ 2010). On the other hand, the redshifted emission of HCO$^+$\,(1--0) is elongated in a direction perpendicular to the HCN outflow (see dotted grey line in Fig.~\ref{fi22198co}). 
Since SiO and HCN outflows are more common in Class 0 protostars than in more evolved Class I objects (\eg\ Tafalla \et\ 2000; Shang \et\ 2006; L\'opez-Sepulcre \et\ 2011; Takahashi \& Ho 2012), the overall picture suggests that there are two outflows in the region, as already proposed by S\'anchez-Monge (2011): outflow A, most likely driven by a Class 0 protostar; and outflow B.
Thus, the CO\,(2--1) knots observed with the PdBI-A could be mainly tracing the cavity walls of outflow B, as most of the knots are seen in pairs along the direction of the HCO$^+$ high-velocity emission, and the strong CO\,(2--1) blueshifted knot $5''$ to the southwest of MM2 (as well as the CO\,(2--1) redshifted emission to the northeast) could be associated with outflow A.
MM2, which is a Class 0-like source, not detected up to 4.5~\mum, and with emission from complex organic molecules (Palau \et\ 2011), is most likely the driving source of outflow A. 
As for the driving source of outflow B, this could be either an unresolved close companion of MM2 (at $\lesssim300$~AU) or MM2-S itself (600~AU to the south of MM2), which falls within $\lesssim1''$ of the only 3.6~\mum\ Spitzer source detected in the field. 
Therefore, in this region there is one strong source dominating the millimeter emission, which is driving at least one outflow.

\paragraph{A5142:} The region has been proposed to harbor a small protocluster, because different millimeter sources (Zhang \et\ 2007), at least three outflows (Hunter \et\ 1995, 1999; Chen \et\ 2005, Zhang \et\ 2007) and multiple water masers (Hunter \et\ 1995, 1999) have been reported. Our PdBI-A observations reveal a striking richness of millimeter sources (especially compared to the other regions), with two strong sources in the center and 5 additional sources all found within $7''$ (12000~AU or 0.06~pc), as well as multiple CO\,(2--1) high-velocity knots (Fig.~\ref{fa5142co}). The strongest CO\,(2--1) redshifted knots are all located along the northwest direction following the well-known outflow C (\eg\ Zhang \et\ 2007), and the fainter redshifted knots seen to the southwest are well aligned with water masers (Goddi \& Moscadelli 2006) and CO emission from outflow B (Zhang \et\ 2007).
Additionally, there are hints of a possible new outflow (knots `D'), which would consist of 3 blueshifted knots well aligned with MM2 and MM6, and extending roughly along the east-west direction. Hints of such an east-west outflow were also found in the blueshifted SO emission by Zhang \et\ (2007, see figure 7 of that paper).
The driving sources of the outflows seem to be all located at the central $2''$ of the protocluster, where MM1 and MM2 are found together with two additional hot cores, MM1-HC and MM2-HC, detected in different complex organic molecules (Fig.~\ref{fa5142co}-right, Palau \et\ 2011). 
As explained in Palau \et\ (2011), MM1-HC is the driving source of outflow C, as its peak falls at the center of symmetry of a thermal radiojet elongated in the southeast-northwest direction (Zhang \et\ 2007; Goddi \et\ 2011), and it is also coincident with the center of symmetry of \chtoh\ masers tracing an infalling disk (Goddi \et\ 2011). Concerning MM1, its millimeter continuum emission elongated in the east-west direction suggests that this source could be driving the north-south outflow A. Outflow B and the blueshifted knots `D' could be related to MM2 and/or MM2-HC, as already proposed by Goddi \& Moscadelli (2006) from water maser observations (note that MM2 is elongated in the north-south direction, thus perpendicular to knots `D', and MM2-HC is elongated perpendicular to outflow B, see Fig.~\ref{fa5142co}-right).
As for the 5 additional millimeter sources surrounding the central part of the protocluster, MM3, MM6 and MM9 are found exactly along outflow axes, and it is not clear if they are tracing dust entrained by the outflows, as recently found by Maury \et\ (2010, 2012), or are true protostellar objects. The fact that MM3 and MM6 are associated with water maser emission at velocities about $\sim5$~\kms\ away from the systemic velocity would rather suggest that they are fragments entrained by the passage of the outflow.
The nature of MM7, close to a 2MASS star, and MM8, not exactly falling along any outflow axis, could be protostellar. In summary, this region presents a rich variety of millimeter sources, with two of them located at the center, dominating the millimeter emission and which seem to drive outflows, and the other fainter millimeter sources being either fragments entrained by the passage of the outflows, or true protostellar deeply embedded low-mass sources.

%%%%%%%%%%%%%%%%%%%%%%%%%%%%%%%%%%%%%%%%%%%%%%%%%%%%%%%%%%%%%%%%%%%%%%%%%%%%%%%%
\section{Analysis}

\subsection{Regions in the literature used to complete the sample: general properties of massive dense cores \label{saprop}}

In order to enlarge our sample, we searched the literature for massive dense cores observed with millimeter interferometers down to spatial resolutions of $\sim1000$~AU and with mass sensitivities $\lesssim0.5$~\mo. 
We required such a spatial resolution and sensitivity because this is the resolution and sensitivity of the optical/infrared studies reporting a correlation between the richness of the cluster and the mass of its most massive star (\eg\ Hillenbrand 1995; Testi \et\ 1999), and our final goal is to study such a correlation in the millimeter/submillimeter range.
The final list, of 18 regions in total, is given in Table~\ref{tstatistics}
\footnote{In our literature search we found two regions, Orion-KL (Beuther \et\ 2004) and G5.89 (Hunter \et\ 2008), which fulfill our selection criteria but were not included for different reasons. Orion-KL is a region undergoing a dynamical interaction between multiple stars (\eg\ Zapata \et\ 2009, 2011; Moeckel \& Goddi 2012), which is a very rare case in the literature; and the 1~mm continuum emission of G5.89 is strongly contaminated by free-free emission (Hunter \et\ 2008).},
where we list the main observational parameters for each region. 
The rms noise of each image, as well as the angular resolution, field of view, and largest angular scale to which the interferometer is sensitive have been converted to mass sensitivity and spatial scales to ease the comparison between regions. 
The last 4 regions of Table~\ref{tstatistics} correspond to 4 regions studied by Bontemps \et\ (2010) which fulfill our selection criteria (Section~1). These 4 regions
are separated from the others because they were observed with the PdBI in CD configuration, recovering spatial structures about a factor 3--5 larger than the other regions of Table~\ref{tstatistics} (see column 9 in Table~\ref{tstatistics}).
A compilation of the millimeter high angular resolution images of all the regions (for the last four see figures 2 and 3 of Bontemps \et\ 2010) are presented in Figs.~\ref{flit1} and \ref{flit2} of the Appendix.

In order to characterize in a uniform way the main properties of the massive dense cores of Table~\ref{tstatistics}, we have analyzed with the same methodology (using the flux densities reported by Motte \et\ (2007) and di Francesco \et\ (2008) for each core, and calculating all the parameters as described below) the following parameters for all the massive dense cores of Table~\ref{tstatistics} (results listed in Tables~\ref{tstatistics} and \ref{tstatistics2}). 

\paragraph{Bolometric luminosity and bolometric temperature (\Lbol, \Tbol)} 
To estimate \Lbol\ and \Tbol\,\footnote{\Tbol\ is defined as the temperature of a blackbody having the same mean frequency as the observed spectral energy distribution (\eg\ Chen \et\ 1995).},  we compiled the flux densities at different frequencies from 2MASS, Spitzer/IRAC+MIPS, WISE, MSX, IRAS, and JCMT (SCUBA, di Francesco \et\ 2008) catalogs. This allowed us to build the spectral energy distribution (SED) of the (presumably) most luminous source embedded in the massive dense core. 
For some regions where Spitzer data were strongly saturated, we used FIR/submillimeter fluxes from the literature 
(Butner \et\ 1990; Campbell \et\ 1995). 
When possible, we compared our compiled photometry (especially the Spitzer/IRAC+MIPS and WISE fluxes) with fluxes reported in the literature at similar wavelengths (\eg\ Lada \et\ 1984; Skinner \et\ 2009; Crimier \et\ 2010), and found them to be consistent.
To build the SED in the infrared we used the flux densities of the 2MASS and Spitzer/IRAC source closest to the strongest millimeter source detected with an interferometer, which fell within $<5''$ of the Spitzer/MIPS and JCMT/SCUBA peaks of emission, for all the cases except for OMC-1S.
\footnote{For the case of OMC-1S, the infrared source chosen to build the SED is source 136$-$355 (Zapata \et\ 2005), which is about $\sim12''$ to the north of the strongest mm source. Given the number and intensity of millimeter sources embedded in the dense core traced by the JCMT at 450~\mum\ (di Francesco \et\ 2008), we estimated that about 1/3 of the total submm flux density at 450 and 850~\mum\ comes from source 136$-$355, and used these estimates to build the SED of this source.}.
Thus, the compiled photometry was derived from data taken with beams $\lesssim20''$, similar to our studied field of view. The obvious exception are the IRAS fluxes, but IRAS, Spitzer/MIPS and WISE were found to be consistent within 30--40\%. This indicates that the IRAS fluxes have their main contribution from the strongest millimeter source of the field and that the effects of multiple sources within the IRAS beam are not dominant in our sample. Overall, we estimated an error of about 30--40\% in the derived bolometric luminosity of all the regions, except for OMC-1S, for which the error should be of a factor of 2 because its SED is the only SED not properly sampled between 3.6~\mum\ and 450~\mum.
Therefore,  we are confident that the \Lbol\ given in Table~\ref{tstatistics} is a reasonable approach to the luminosity (and hence to the mass) of the most massive star embedded in each massive core.
Concerning \Tbol, our derived values span a range of one order of magnitude, from 30 to $\sim200$~K, with 10 cores having \Tbol$\lesssim60$~K.

\begin{figure}
\begin{center}
\begin{tabular}[b]{c}
    \epsfig{file=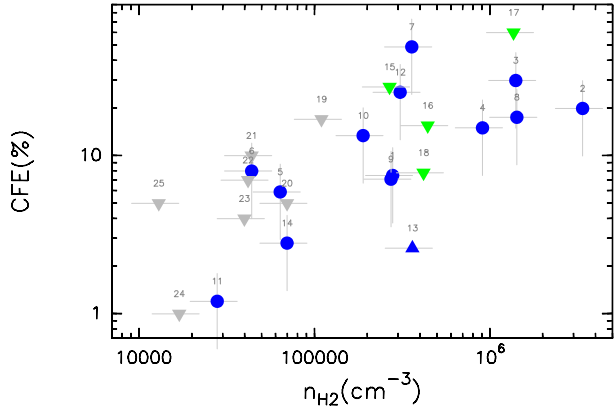, width=8cm, angle=0}\\
\end{tabular}
\caption{CFE vs mean density of the core, $n_\mathrm{H_2}$, after Bontemps \et\ (2010). Green triangles: Cygnus\,X cores (Bontemps \et\ 2010); grey triangles: $\rho$-Oph cores (Motte \et\ 1998); blue dots: cores presented in this work. For Bontemps' and Motte's cores, the CFE is an upper limit (compared to the other cores) because the values are measured with a compact configuration of an interferometer.
}
\label{fCFE_n}
\end{center}
\end{figure}

\paragraph{Mass, mean surface density, and mean density of the core ($M_\mathrm{SD}$, $\Sigma$, $n_\mathrm{H_2}$)} The mass of each dense core was estimated using the JCMT/SCUBA flux densities at 850~\mum\ (di Francesco \et\ 2008, Jenness \et\ 1995), or the 1.2~mm flux densities for the Cygnus\,X regions (Motte \et\ 2007), adopting the new distance to Cygnus\,X of 1.4~kpc (Rygl \et\ 2012).
To estimate the mass we assumed a dust temperature of 20~K, and a dust (+gas) mass opacity coefficient of 0.0175\,cm$^2$\,g$^{-1}$ at 850~\mum\ and 0.00899\,cm$^2$\,g$^{-1}$ at 1.2~mm (Ossenkopf \& Henning 1994). 
The masses range from 17 to 2500~\mo, and increase with \Lbol\ (in Table~\ref{tstatistics2}, the regions are ordered with increasing luminosity), as already seen in previous works (\eg\ Sridharan \et\ 2002). By using the FWHM of the 850~\mum\ core (measured in the images of di Francesco \et\ 2008, and reported in Motte \et\ 2007), and assuming spherical symmetry, we derived the surface density and mean density\footnote{We estimated the surface density following $\Sigma=M/(\pi\,R^2)$ (with $R=FWHM/2$) and the mean density following $n=3\,M/(4\pi\,FWHM)$, as in Motte \et\ (2007) and Bontemps \et\ (2010).} of each core. These parameters range from 0.7 to 10~g\,cm$^{-2}$, and from 0.3 to 30 times 10$^5$~\cmt. 

\paragraph{Ratio of bolometric luminosity to mass of the core (\Lbol/$M_\mathrm{SD}$)} \Lbol/$M_\mathrm{SD}$ was estimated as an alternate measure (together with \Tbol) of the evolutionary stage of the dense core (\eg\ Sridharan \et\ 2002; L\'opez-Sepulcre \et\ 2011). While \Lbol/$M_\mathrm{SD}$ is more sensitive to $M_\mathrm{SD}$ and thus more representative of the dense core as a whole, \Tbol\ is more sensitive to the properties of the infrared source chosen to build the SED, as \Tbol\ assesses the mean frequency of the SED weighted with the flux density at each frequency.
In the following, we will use \Lbol/$M_\mathrm{SD}$ as a proxy to the core evolutionary stage.

\paragraph{Core Formation Efficiency (CFE)} We calculated the CFE as the ratio between the sum of the masses of all the fragments measured with an interferometer in very extended configuration (Table~\ref{tstatistics}) and the single-dish mass estimated from the peak intensity. We used the single-dish peak intensity because the interferometer field of view, $\sim20''$, is similar to the single-dish beam ($\sim23''$ at 850~\mum, di Francesco \et\ 2008). The values of CFE range from 0.4 to 60\%.
Interestingly, we find a trend (Fig.~\ref{fCFE_n}) between CFE and the mean density of the core, which fits within the trend previously found by Motte \et\ (1998) and Bontemps \et\ (2010; see grey and green dots in Fig.~\ref{fCFE_n}), and which is suggestive of gravity playing a dominant role in the densest cores, favoring the formation of compact structures.

\begin{figure}
\begin{center}
\begin{tabular}[b]{c}
    \epsfig{file=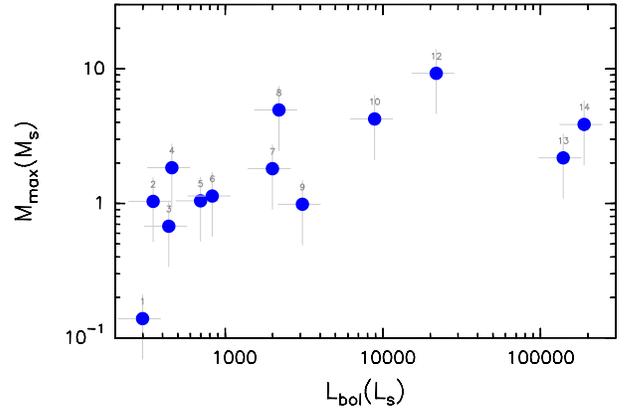, width=8cm, angle=0}\\
\end{tabular}
\caption{$M_\mathrm{max}$ vs \Lbol\ for the (sub)sample of 14 sources. The figure shows a possible weak trend between  $M_\mathrm{max}$ (the mass of the most massive fragment as measured with an interferometer in extended configuration) and \Lbol. In this plot I22134 is not included because the two most massive sources of this region (the UCHII region and IRS1), are almost completely deprived of dust (Section~\ref{srco}), and cannot be compared to figure 2 of Bonnell \et\ (2004), calculated at the time of protostar formation (see Section~\ref{sdmsf}).
}
\label{fMmax_Lbol}
\end{center}
\end{figure}

\paragraph{Mass of the strongest millimeter source ($M_\mathrm{max}$)} $M_\mathrm{max}$ was calculated from the flux density of the strongest millimeter source measured by an interferometer with a very extended configuration (see Table~\ref{tstatistics} for the references; the Cygnus\,X sources have been excluded as they were observed with the PdBI in CD configuration), assuming a dust temperature of 50~K and the dust opacity law of Ossenkopf \& Henning (1994) as in Table~\ref{tcont}. 
Some of these sources are associated with hot molecular cores for which temperatures of $\sim100$~K have been estimated. However, this information is not available for all the sources, and for the temperature range of 30--100~K the masses change by a factor of 2 with respect to the value derived assuming 50~K, which we consider a reasonable first approach.  
The plot of $M_\mathrm{max}$ vs \Lbol\ (Fig.~\ref{fMmax_Lbol}) shows a weak correlation, {which will be further discussed in Section~\ref{sdmsf}}.

\begin{figure*}
\begin{center}
\begin{tabular}[b]{c}
    \epsfig{file=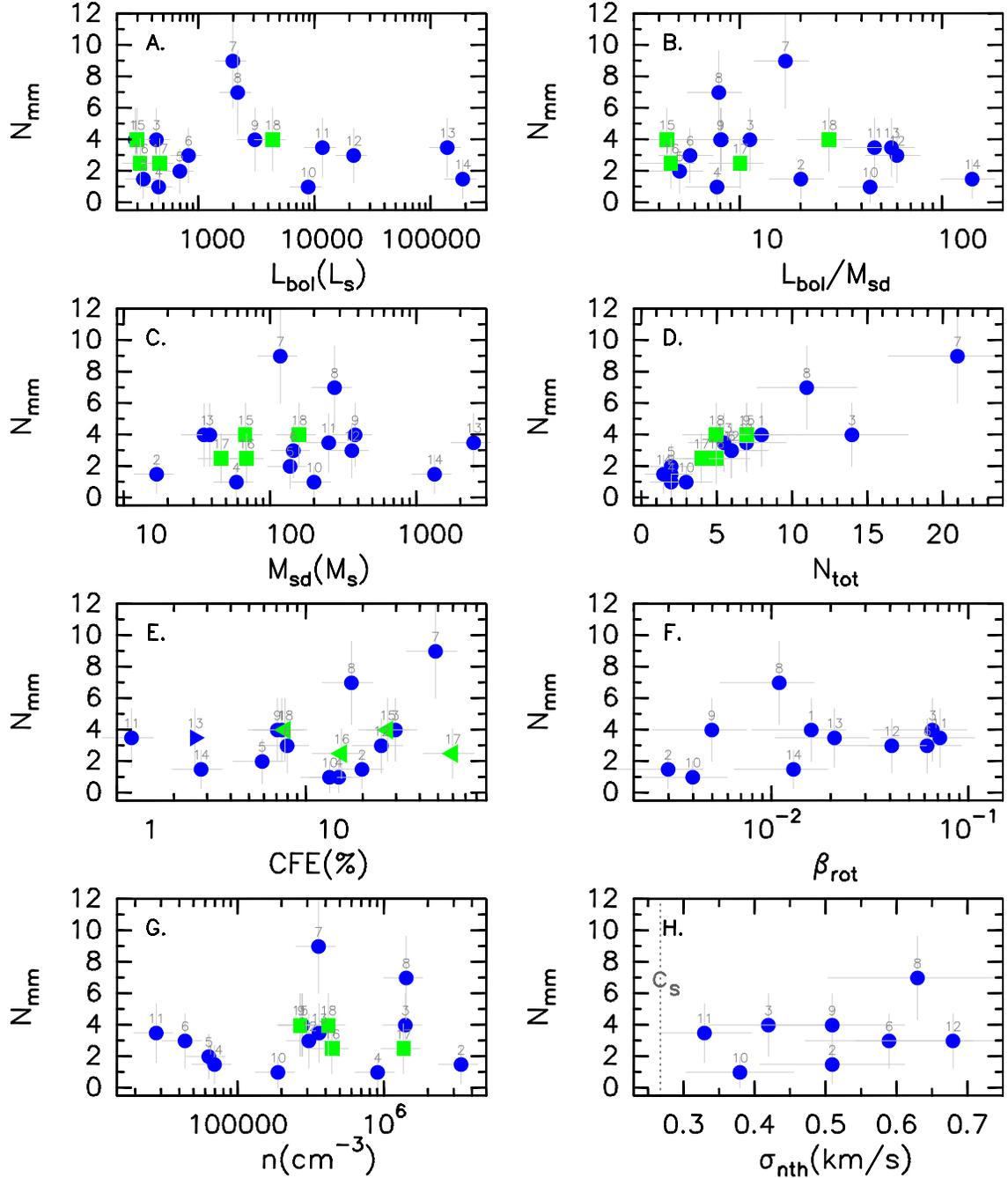, width=15cm, angle=0}\\
\end{tabular}
\caption{\Nmm\ vs a number of properties etimated for the 18 massive dense cores and compiled in Tables~\ref{tstatistics} and \ref{tstatistics2} (see Section~\ref{saprop}). The numbers on the top of each dot indicate the source-ID listed in Tables~\ref{tstatistics} and \ref{tstatistics2}. Blue dots correspond to sources labeled from 1 to 14 in the tables, and the green dots correspond to sources 15 to 18, which are the Cygnus\,X sources studied in Bontemps \et\ (2010). In panel H, the vertical dotted line marks the sound speed ($c_\mathrm{s}$, corresponding to a temperature of 20~K).
}
\label{f8Nmm}
\end{center}
\end{figure*}

\paragraph{Non-thermal velocity dispersion ($\sigma_\mathrm{non-th}$)} To have a first idea of the initial non-thermal velocity dispersion in the core, we searched for interferometric data of \nh(1,1) reported in the literature and found data for 8 regions (Zhou \et\ 1990; Tieftrunk \et\ 1998; Zhang \et\ 1998; Zhang \et\ 2002; G\'omez \et\ 2003; Busquet 2010; Keto \& Zhang 2010; S\'anchez-Monge 2011; S\'anchez-Monge \et\ in prep.; all observed with the Very Large Array with a typical angular resolution of $\sim5''$ and field of view $\sim2'$). Among the different \nh\ cores surrounding the massive dense core of our sample, we chose the most quiescent core, which should be barely affected by turbulence injected by outflows or by systematic motions, and should be representative of the initial turbulence of the massive dense core. With the measured linewidth in this quiescent core, we estimated the non-thermal velocity dispersion, $\sigma_\mathrm{non-th}$ (assuming a thermal component corresponding to 20~K), and found values from 0.3 to 0.7~\kms\,\footnote{For the Cygnus\,X sources, Bontemps \et\ (2010) estimate the non-thermal velocity dispersion from single-dish measurements at the position of each massive core, with an angular resolution of $\sim20''$, thus possibly including turbulence injected by outflows/systematic motions, and for this reason we do not include these values here.}, all supersonic (sound speed corresponding to 20~K is 0.27~\kms).
The narrow range of values estimated for $\sigma_\mathrm{non-th}$, around 0.5~\kms, suggests that our approach is reasonably avoiding outflows/infall/rotation motions.

\paragraph{Jeans mass and Jeans number ($M_\mathrm{Jeans}$, $N_\mathrm{Jeans}$)} 
We estimated the `mean' Jeans mass using the mean density of the cores listed in Table~\ref{tstatistics2}, and assuming a gas temperature (as in gravo-turbulent fragmentation scenarios, \eg\ Schmeja \& Klessen 2004; Dobbs \et\ 2005), and found $M_\mathrm{Jeans}\sim$0.5--3~\mo\ (Table~\ref{tstatistics2}). This is much smaller than the mass of the core, and larger than the mass sensitivity achieved for regions of our sample (Table~\ref{tstatistics}), implying that we should have detected between 20 and 200 fragments if a pure thermal Jeans fragmentation was at play. We also calculated $M_\mathrm{Jeans}$ using the total velocity dispersion (thermal+non-thermal, also called 'turbulent Jeans' fragmentation, Pillai \et\ 2011), yielding $M_\mathrm{Jeans}\lesssim~M_\mathrm{SD}$, which gives probably an idea of the mass of the most massive fragments expected. Thus, according to Jeans fragmentation, one would expect a large number of low-mass fragments and a few massive fragments, while we observe only a few low-mass fragments.

\paragraph{Rotational-to-gravitational energy ($\beta_\mathrm{rot}$)} 
Simulations show that rotating cores fragment if their rotational energy is large enough compared to the gravitational energy (\eg\ Boss 1999; Forgan \& Rice 2012). 
The threshold derived from the simulations is $\beta_\mathrm{rot}>0.01$, and recent observations of a sample of low-mass rotating systems seem to be consistent with this lower limit (Chen \et\ 2012).
In our sample, $\beta_\mathrm{rot}$ was estimated following equation (1) of Chen \et\ (2012), and using interferometric observations of  \nh(1,1) (for 8 regions, see above), \nh(2,2), CS\,(2--1), and SO$_2$ (Torrelles \et\ 1989; Beltr\'an \et\ 2004; Rod\'on \et\ 2008),
tracing the core at scales of 5--$10''$. We caution that the velocity gradients seen in the dense cores, used to estimate $\beta_\mathrm{rot}$, could be affected in some cases by multiple velocity components and/or perturbation by outflows.
Despite of this, the estimated values of $\beta_\mathrm{rot}$ range from 0.003 to 0.072 (Table~\ref{tstatistics2}), similar to the range of values reported in the literature (\eg\ Goodman \et\ 1993; Chen \et\ 2012).

\subsection{Fragmentation level in massive dense cores \label{saNmm}}

The total sample of 18 massive dense cores compiled in Table~\ref{tstatistics} spans a range of bolometric luminosities from $\sim300$ up to $2\times10^5$~\lo, or almost 3 orders of magnitude, and we aim at relating the dense core properties, such as \Lbol, with their fragmentation level. To achieve this, we did the following.
For each high angular resolution millimeter image (see Table~\ref{tstatistics} for references) we estimated the rms noise using the same procedure in \textsc{Mapping} of the \textsc{Gildas} software and identified those sources above 6$\sigma$ (in Section~\ref{srcont} we showed that the 6$\sigma$ cutoff yields reliable identifications; the 6$\sigma$ level is marked in red in the cleaned images of Figs.~\ref{flit1} and \ref{flit2}). We defined \Nmm\ as the number of millimeter sources inside a region of 0.1~pc of diameter (marked in the figures of the Appendix as a black circle). In case of 4 or 5$\sigma$ detection but with independent claims in the literature of being a true source, we counted the source as `half' (0.5). 
The median number of \Nmm\ in the sample is 3.25.

Among the 18 studied cores in Table~\ref{tstatistics}, five ($\sim28$\%) show a very low degree of fragmentation, with only one source (or one dominant source with a much fainter companion) within the 0.1~pc region, while 9 ($\sim50$\%) cores split up into $\gtrsim4$ sources.
The sources with a low degree of fragmentation are prototypical deeply embedded intermediate/high-mass sources  (\eg\ Cesaroni \et\ 2005; van der Tak \et\ 2006; Fuente \et\ 2005; S\'anchez-Monge \et\ 2010; Bontemps \et\ 2010), as their low \Tbol\ and \Lbol/$M_\mathrm{SD}$ indicate (see Tables~\ref{tstatistics} and \ref{tstatistics2}). Therefore, the observed low fragmentation is not an evolutionary effect of clusters which have already cleared out their natal material.

In addition, in order to have a first estimate of the infrared sources possibly associated with each massive dense core, we searched the 2MASS Point Source Catalogue (Skrutskie \et\ 2006) and downloaded the images from Spitzer/IRAC database\footnote{Data downloaded from the 2MASS and Spitzer Heritage Archive webs, http://www.ipac.caltech.edu/2mass/ and http://sha.ipac.caltech.edu/applications/Spitzer/SHA.} to extract the positions of the 3.6~\mum\ sources within the 0.1~pc field of view for each core. The final positions of the infrared (2MASS and 3.6~\mum) sources are marked in Figs.~\ref{flit1} and \ref{flit2}.
We counted the number of infrared sources (2MASS + Spitzer/IRAC at 3.6~\mum) within 0.1~pc (see Table~\ref{tstatistics}) and found that the total number of millimeter and infrared sources, $N_\mathrm{IR+mm}$, correlates with \Nmm, \ie\ the regions with a small number of millimeter sources also seem to have a small number of infrared sources, and vice versa (see Table~\ref{tstatistics} and Fig.~\ref{f8Nmm}-d). Similarly, the ratio $N_\mathrm{IR}$/$N_\mathrm{mm}$ is in general close to 1 (Table~\ref{tstatistics}).
We are aware that there are several factors contributing to the uncertainty of $N_\mathrm{IR}$. On the one hand, the mass sensitivity of 2MASS is not as good as the sensitivity of the PdBI-A, especially if there is important visual extinction in the region (as is the case for our regions). We estimated that, following the method of Testi \et\ (1998), for an (optimistic) visual extinction of $A_\mathrm{V}\sim5$~mag, the mass sensitivity of 2MASS in $K_\mathrm{s}$-band (which has a limiting magnitude of 14.3, Skrutskie \et\ 2006) is 0.6--1.4~\mo, for an age of 0.5--1~Myr. On the other hand, we did not do any color-color analysis to discard possible foreground/background contaminants.
As for the Spitzer/IRAC 3.6~\mum\ data, for a region at 2~kpc, we are sensitive to $\sim0.1$~\mo\ (using a limiting magnitude of $\sim16.5$, Qiu \et\ 2008). The only caveat with Spitzer/IRAC images is that the two brightest sources of the sample (W3IRS5 and AFGL\,2591) are affected by saturation effects, and thus $N_\mathrm{IR}$ is a lower limit in these two cases. 
In spite of these shortcomings, it is striking that the ratio $N_\mathrm{IR}$/$N_\mathrm{mm}$ is $\sim1$ for 12 out of 18 sources and, although the exact number of infrared sources embedded in the massive dense cores should be studied more accurately, the trend shown in Fig.~\ref{f8Nmm}-d is probably real.
Overall, the total number of millimeter sources found in all the regions is 57, of which only 20 have infrared counterparts. This indicates that only $\sim30$\% of the detected millimeter sources are also detected in the infrared, consistent with previous studies by Qiu \et\ (2008), and indicating that in general the millimeter sources in our sample of massive dense cores are deeply embedded and in very early evolutionary stages.

As a last calculation, we assessed the protostellar density by dividing the total number of millimeter+infrared sources (Table~\ref{tstatistics}) by the volume corresponding to a sphere 
with the same projected area onto the sky as the polygon with minimum area encompassing all the sources.
The result is shown in Table~\ref{tstatistics}, togethter with the mean separation between sources, typically of $\sim3000$~AU.
These numbers should be regarded as preliminary and as lower limits for W3IRS5 and AFGL\,2591 due to limitations of the infrared datasets (see above). The average stellar density is around $\sim10^5$\,stars\,pc$^{-3}$,
similar to the stellar density measured in the Orion Nebula Cluster (\eg\ Hillenbrand \& Hartmann 1998).

%%%%%%%%%%%%%%%%%%%%%%%%%%%%%%%%%%%%%%%%%%%%%%%%%%%%%%%%%%%%%%%%%%%%%%%%%%%%%%%%
\section{Discussion}

\subsection{Implications for intermediate/high-mass star formation \label{sdmsf}}

In Section~\ref{saprop}, we found a weak trend between the mass of the strongest millimeter source and the bolometric luminosity estimated for each core (Fig.~\ref{fMmax_Lbol}).
In the original competitive accretion scenario, Bonnell \et\ (2004) predicted that  the mass of the small-scale (1000--2000~AU) condensation associated with the nascent intermediate/high-mass star does not correlate with the final stellar mass at the time of protostar formation
(see figure~2 of Bonnell \et\ 2004). The weak trend shown in Fig.~\ref{fMmax_Lbol} does not support a pure competitive accretion scenario.
On the other hand, Fig.~\ref{fCFE_n} shows that for most of the cores only a small fraction of mass is found in compact fragments, which means that most of the mass is in an extended envelope outside the fragments. This finding counters the monolithic collapse scenario (McKee \& Tan 2002, 2003). Thus, the results presented here (Figs.~\ref{fCFE_n} and \ref{fMmax_Lbol}) show that neither monolithic collapse nor competitive accretion accurately prescribe the physics in these regions. Zhang \& Wang (2011) propose an alternative picture that condensations/fragments continue to accrete mass from the general environment (called `core' in this paper), whereas the embedded protostar accretes mass from the condensation/fragment. This `simultaneous accretion picture' explains both Figs.~\ref{fCFE_n} and \ref{fMmax_Lbol}.

\subsection{What is controlling fragmentation in massive dense cores? \label{sdfrag}}

In the previous section we presented a sample of 18 massive dense cores observed in the millimeter range down to a spatial resolution of $\sim1000$~AU and mass sensitivities $\lesssim0.5$~\mo. Among the 18 cores, about $\sim30$\% show no fragmentation, while 50\% harbor $\gtrsim4$ fragments. Given the different fragmentation levels in our sample, we consider in the following possible ingredients which could play a role in the fragmentation of a core. 

Optical and infrared studies suggest that \Lbol\ is related to the number of sources in a cluster (\eg\ Hillenbrand 1995; Testi \et\ 1998). 
We searched for any correlation between \Nmm\ and \Lbol, and the result is plotted in Fig.~\ref{f8Nmm}-a. We find no signs of any possible correlation: the highest values of \Nmm\ are reached in regions of low/intermediate luminosity, such as OMC-1S and A5142, and regions with similar luminosity show very different properties (compare for example NGC\,7129-FIRS2, Fuente et al. 2005, to IC1396-N, Neri et al. 2007). 
Since \Lbol\ correlates well with $M_\mathrm{SD}$ (see Section~\ref{saprop}), this could indicate that the mass of the core is not a crucial ingredient in determining its fragmentation (see also Fig.~\ref{f8Nmm}-c). This could also indicate that the correlation seen in the optical/infrared is probably not established at early stages of cluster formation, and that the final number of members in a cluster is set after several star formation episodes, as suggested by both theoretical (\eg\ Myers 2011) and observational (\eg\ Williams \et\ 2009; Hennemann \et\ 2010; Bik \et\ 2012) works. A deeper study including a larger sample and larger field of view should be carried out to draw firm conclusions on this. We note however that cluster evolution alone is not likely enough to explain the different fragmentation, since for cores with a similar value of the evolutionary indicators ($T_\mathrm{bol}$, \Lbol/$M_\mathrm{SD}$) we already find very different fragmentation levels (see Fig.~\ref{f8Nmm}-b).
This was in fact also suggested by the trend found between $N_\mathrm{mm}$ and $N_\mathrm{mm+IR}$ (Fig.~\ref{f8Nmm}-d; or $N_\mathrm{IR}$/$N_\mathrm{mm}\sim1$, Table~\ref{tstatistics}) indicating that the different episodes of star formation yield comparable numbers of fragments, with some cores highly fragmenting from the beginning and keeping this property constant with time, while others remaining unfragmented during all core evolution.

The initial density profile of a core could be one of the agents determining its fragmentation level, as simulations show that highly concentrated cores (\eg\ $n_\mathrm{H_2}\propto r^{-2}$) lead to the formation of one single massive star, while flat density profiles  ($n_\mathrm{H_2}\propto r^{-1}$ or flatter)  produce an important number of low-mass stars (\eg\ Myhill \& Kaula 1992; Burkert \et\ 1997; Girichidis \et\ 2011). In our sample, two of the 5 cores showing no fragmentation present concentrated density profiles (Beuther \et\ 2002a; Mueller \et\ 2002; Crimier \et\ 2010), supporting this view. A uniform study of the density profiles of the cores presented in this work would be required to accurately assess its importance on the fragmentation level.

Another ingredient which could be related to fragmentation is the rotational-to-gravitational energy, $\beta_\mathrm{rot}$. The calculated values of $\beta_\mathrm{rot}$ (Sec.~\ref{saprop}, Table~\ref{tstatistics2}) present no relation with \Nmm\ (Fig.~\ref{f8Nmm}-f), although on average, the low-fragmented regions have $\beta_\mathrm{rot}$ smaller than the highly fragmented regions. The highly fragmented regions present a wide range of $\beta_\mathrm{rot}$ values spannig at least one order of magnitude. This could indicate that $\beta_\mathrm{rot}$ is not an accurate measure of the rotation of the cores (\ie\ the velocity gradients measured may not always trace rotation), or that a wide range of angular momenta can yield similar fragmentation.

The turbulence associated with a core could be also determining its fragmentation. For example, `turbulent fragmentation' models show that turbulence develops shocks which in turn produce substructure in the core (\eg\ Padoan \& Nordlund 2002; Klessen \et\ 2005), and one could expect a higher fragmentation for higher turbulence levels (Jappsen \& Klessen 2004; Peretto \et\ 2007). However, \Nmm\ does not seem to depend either on the non-thermal (turbulent) velocity dispersion of the cores of our sample (Section~\ref{saprop}, Fig.~\ref{f8Nmm}-h). 

Finally, in Section~\ref{saprop} we showed that there is a deficit in the observed number of low-mass fragments as compared to expectations from thermal Jeans fragmentation, and this is true even for cores with rather flat density profiles (\eg\ I22134, IRAS\,05358+3543NE, Beuther \et\ 2002a).
This is very similar to the conclusions drawn by Bontemps \et\ (2010) and van Kempen \et\  (2012) for their (samples of) sources.

All this suggests that still other forms of support additional to thermal and turbulence may play a role. 
Interferometric observations of polarized submillimeter emission have shown that the magnetic field seems to play an important role in the dynamics of some massive dense cores  (\eg\ Lai \et\ 2003; Girart \et\ 2009; Tang \et\ 2009).
In the following section we discuss the influence of magnetic field and radiative feedback on the fragmentation process of a massive dense core.

\begin{figure}
\begin{center}
\begin{tabular}[b]{c}
    \epsfig{file=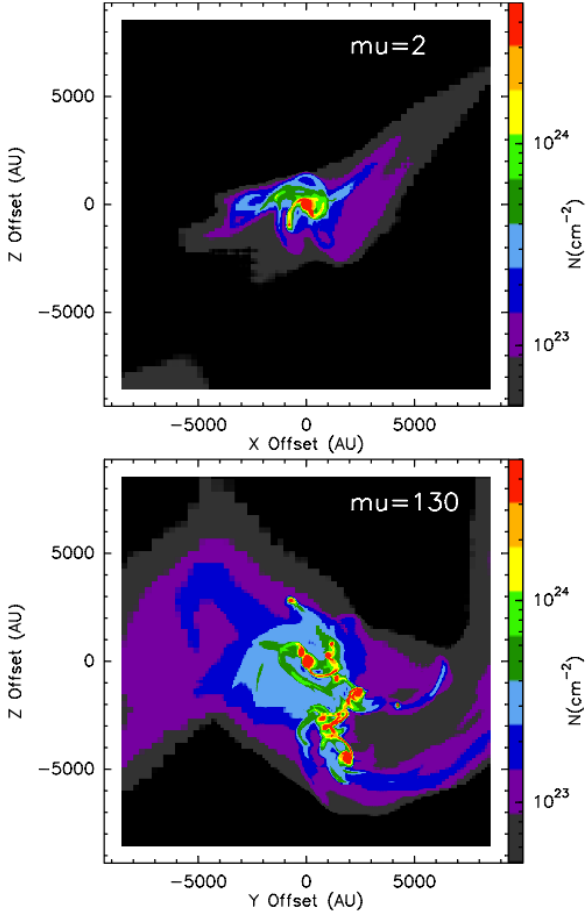, width=8.cm, angle=0}\\
\end{tabular}
 \caption{Direct output of Commer\c con \et\ (2011a) simulations (column density maps) for the highly magnetized core model (top) and the low magnetized core model (bottom).
}
\label{fsimraw}
\end{center}
\end{figure}

\subsection{Role of combined magnetic field, radiative feedback and turbulence in the fragmentation process}

A number of studies comparing observations to simulations conclude that magnetic fields could be a key ingredient in the fragmentation process. For example, for the low-mass case, Maury et al. (2010) find that magneto-hydrodynamical models agree much better with their observations. And for the intermediate/high-mass case, Peretto \et\ (2007) find difficulties matching the observed masses and number of fragments with the results of hydrodynamical simulations, suggesting that an extra support such as protostellar feedback or magnetic fields is at play.
In order to assess the role of magnetic fields and radiative feedback in the fragmentation of a massive core we considered the radiation magneto-hydrodynamic simulations of Commer\c con \et\ (2011a). These simulations include the effects of magnetic fields, turbulence, and take into account the radiative feedback from the accretion shock (i.e. shock on the first Larson core) and thermal emission (no radiative feedback from the accreting/forming protostars using pre-main sequence stellar evolution subgrid models), down to a spatial resolution of 32~AU (low resolution runs, see fig. 2 in Commer\c con \et\ 2011a). 
We note that the initial conditions adopted by Commer\c con \et\ (2011a) correspond to a relatively flat density profile, which should favor fragmentation (\eg\ Girichidis et al. 2011).
Commer\c con \et\  (2011a) find that magnetic fields and radiation can combine to inhibit fragmentation in a massive core, and that this is particularly important if the initial magnetic intensity is strong. In their simulations, the initial magnetic intensity is set by the parameter $\mu=(M/\Phi)/(M/\Phi)_\mathrm{crit}$ (mass-to-flux over critical mass-to-flux ratio, Mouschovias \& Spitzer 1976; Hennebelle \et\ 2011), and the two extreme cases explored are $\mu=2$ (highly magnetized core) and $\mu=130$ (weakly magnetized core). By using the output of the adaptive mesh refinement code {\ttfamily RAMSES} (Teyssier 2002; Fromang \et\ 2006; Commer{\c c}on \et\ 2011b), we compute column density maps (in the three different $xy$, $xz$, and $yz$ projections) for each value of $\mu$ from the low resolution runs of Commer\c con \et\ (2011a).

\begin{table}
\caption{Comparison of Commer\c con \et\ (2011a) simulations to PdBI observations of this work within a field of view of $\sim0.07$~pc}
\begin{center}
{\small
\begin{tabular}{lcc}
\noalign{\smallskip}
\hline\noalign{\smallskip}
Parameter
&Simulations
&Observations
\\
\hline\noalign{\smallskip}
\hline\noalign{\smallskip}
\multicolumn{3}{c}{\emph{MU2 - highly magnetized core}}\\
\hline\noalign{\smallskip}
$M_\mathrm{5000\,AU}^\mathrm{tot}$ (\mo)\supa\         	&4.7       		&3.2\\
$M_\mathrm{500\,AU}$ (\mo)\supa\            			&3.8     		&1.9\\
CFE\supb\ 									&81\%		&59\%\\
number of fragments							&1			&1\\
\hline\noalign{\smallskip}
\hline\noalign{\smallskip}
\multicolumn{3}{c}{\emph{MU130 - weakly magnetized core}}\\
\hline\noalign{\smallskip}
$M_\mathrm{5000\,AU}^\mathrm{tot}$ (\mo)\supa\ 		&16       			&47\\
$M_\mathrm{500\,AU}$ (\mo)\supa\ 				&4.0,3.9,1.7		&8.9,6.4,0.94,\\
											&0.77,0.47		&0.85,0.81,0.77,0.72\\
CFE\supb\ 									&68\%			&41\%\\
number of fragments							&5				&7\\
fragment separation (AU)							&2000--3500		&2000--5000\\
\hline
\end{tabular}
\begin{list}{}{}
\item[$^\mathrm{a}$] Masses estimated assuming a dust temperature of 30 K for all the fragments, and the same opacity as in Table~\ref{tcont}. $M_\mathrm{500\,AU}$ refers to the mass of each fragment as observed with the PdBI in A configuration.
\item[$^\mathrm{b}$] Core Formation Efficiency, defined as the ratio between the total mass measured with the PdBI in A configuration ($\sim500$~AU scale) to the mass measured with an interferometer in a compact configuration ($\sim5000$~AU scales). Note that this definition is different from the one used in Section~\ref{saprop}, Table~\ref{tstatistics2}, and Figs.~\ref{fCFE_n} and \ref{f8Nmm}.
\end{list}
}
\end{center}
\label{tsimmul}
\end{table}

\begin{figure*}
\begin{center}
\begin{tabular}[b]{c}
    \epsfig{file=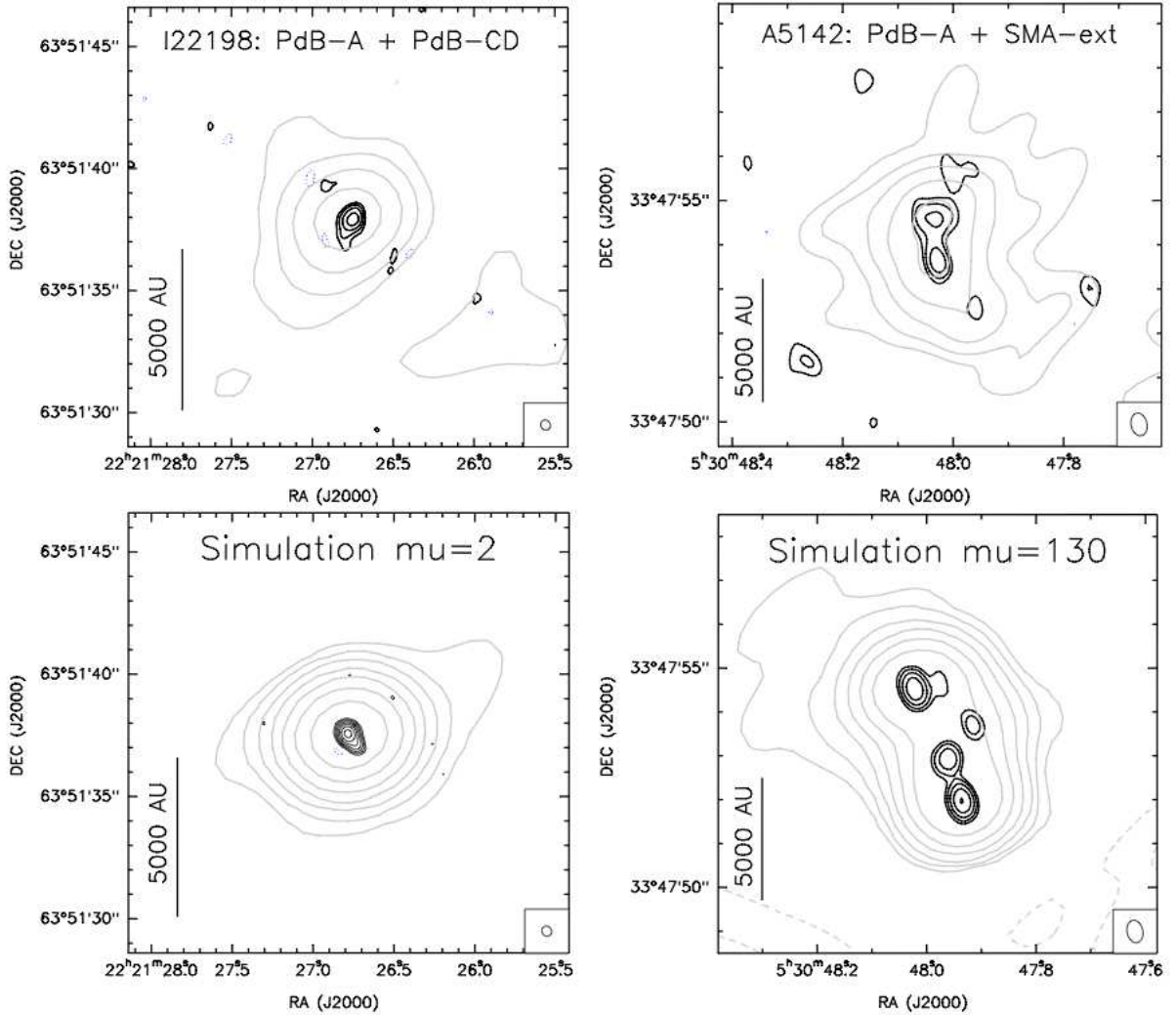, width=16cm, angle=0}\\
\end{tabular}
 \caption{        
         {\bf Top:} Observations of I22198 (left) and A5142 (right) with the PdBI at $0.4''$ angular resolution (black contours) and the PdBI/SMA at $1''$--2$''$ (grey contours; left: I22198 with the PdBI at 3~mm, S\'anchez-Monge 2011; right: A5142 with the SMA at 1.3~mm, Zhang \et\ 2007). 
The field of view corresponds to the same spatial scale for both regions (0.07~pc). 
{\bf Bottom:} Simulations convolved (see footnote~21) with the $uv$-coverage of the PdBI-A (black contours) and PdBI-CD (grey contours). Bottom-left panel are the simulations for the magnetized core case ($\mu=2$). Bottom-right panel are the simulations for the low magnetization case ($\mu=130$). 
Left panels contours: Black: $-4$, 4, 8, 16, 32, 64, 128, and 256 times 2(1)~m\jpb\ for observations (simulations); grey contours: $-4$, 4, 8, 16, 32, 64, 128, and 256 times 0.6(1)~m\jpb\ for observations at 3~mm (simulations at 1~mm).
Right panels contours: Black: $-4$,  4, 8, 16, 32, and 64 times 2.8(1.4)~m\jpb\ for observations (simulations); grey contours: $-2$, $-1$, 1, 2, 4, 8, 16, and 32 times 12(1.2)~m\jpb\ for observations (simulations).
}
\label{fsimPdB}
\end{center}
\end{figure*}

In order to compare the output of the simulations of Commer\c con \et\ (2011a) with our observational interferometric data we selected two regions with low and high fragmentation as seen with the PdBI-A. For the low fragmentation case we selected I22198 to be compared to the $\mu=2$ (highly magnetized) simulated maps, while for the high fragmentation case we selected A5142 to be compared to the $\mu=130$ (weakly magnetized) simulated maps. The column density simulated maps used for the comparison are shown in Fig.~\ref{fsimraw}, which reveal a striking intrinsic difference between the highly magnetized ($\mu=2$) and the weakly magnetized ($\mu=130$) case.
By using the output of the adaptive mesh refinement (AMR) code {\ttfamily RAMSES} (Teyssier 2002; Fromang \et\ 2006; Commer{\c c}on \et\ 2011b), we compute dust emission maps using the {\ttfamily RADMC-3D} code\footnote{http://www.ita.uni-heidelberg.de/$\sim$dullemond/software/radmc-3d/} (in the three different $xy$, $xz$, and $yz$ projections) for each value of $\mu$ from the low resolution runs of Commer\c con \et\ (2011a). For the conversion to Jy\,px$^{-1}$, the 3D AMR structure of {\ttfamily RAMSES} is directly loaded into {\ttfamily RADMC-3D} and we assume that the gas temperature computed in the radiation-magnetohydrodynamics calculations of Commmer\c con \et\ (2011a) is correct and equals the dust temperature. We use the opacity law of Ossenkopf \& Henning (1994) at 1.3~mm as in Sec.~\ref{srcont}. We assume the corresponding distances for I22198 and A5142, and put the coordinates of I22198 to the $\mu=2$ map, and the coordinates of A5142 to the $\mu=130$ map. Finally, we convolved the resulting simulated maps in Jy\,px$^{-1}$ by the $uv$-coverage of PdBI-A used in this work\footnote{For the convolution of the simulated maps by the $uv$-coverage of PdBI-A, we used the GILDAS task `uv\_model', which generates a set of visibilities at the $uv$-baselines of the PdBI-A observations (by calculating the Fourier transform of the simulated image), allowing us to mimick not only the synthesized beam of our observations, but also the effects of filtering large-scale structures by the interferometer.}, inverted the visibilities and cleaned the dirty simulated maps as done with our real observations, allowing us to compare the fragments, at $\sim1000$~AU scales, that the PdBI-A would have detected if our regions were similar to the simulated ones. The results are shown in Fig.~\ref{fsimPdB} (black contours; left panels of the figure for I22198, and right panels for A5142). In order to compare also the morphology of the observed massive dense core with the simulated one at scales $\sim5000$~AU, we convolved the simulated Jy\,px$^{-1}$ images by the the $uv$-coverage of PdBI in a typical CD configuration of the PdBI (grey contours of Fig.~\ref{fsimPdB}-bottom).

As shown in Fig.~\ref{fsimPdB}-left-bottom, for the highly magnetized core ($\mu=2$) the simulations convolved with the PdBI-A $uv$-coverage reveal one single fragment at the center, which is dominating the millimeter emission and which is slightly resolved. This is very similar to what we observe in I22198 (Fig.~\ref{fsimPdB}-left-top), where one single slightly elongated fragment dominates the millimeter emission down to a mass sensitivity of $\sim0.1$~\mo.
Concerning the weakly magnetized core ($\mu=130$), Fig.~\ref{fsimPdB}-right-bottom shows that the core splits up into five fragments, with 3 of them dominating the millimeter emission. This is again similar to what we observe in A5142 (Fig.~\ref{fsimPdB}-right-top), where the large-scale envelope (Zhang \et\ 2007) splits up into 6--7 fragments for a mass sensitivity of $\sim1$~\mo, with two massive fragments in the center.
The qualitative agreement between our PdBI observations and the simulations of Commer\c con \et\ (2011a) is remarkably good.

In Table~\ref{tsimmul} we present a preliminary quantitative comparison of our observations and the simulations of Commer\c con \et\ (2011a). The table compares the mass of the core at $\sim5000$~AU scales to the mass predicted by the simulations, as well as the number of fragments, mass of the fragments, CFE\footnote{The Core Formation Efficiency (CFE) is measured here as the total mass in fragments (at $\sim500$~AU scales) versus the mass of the core at $\sim5000$~AU.}, and average separation between fragments. Again, the comparison between Commer\c con \et\ (2011a) simulations and our observations are consistent. For the $\mu=130$ case the simulated mass of the dense core at scales of $\sim5000$~AU is lower by around a factor of 3 with respect to the observations, but this could be the consequence of the $uv$-coverage used to reproduce the large-scale core (grey contours in Fig.~\ref{fsimPdB}), which was not exactly the same as the $uv$-coverage of the A5142 large-scale observations (taken with the Submillimeter Array, Zhang \et\ 2007). In addition, the fact that the calculations are stopped at an arbitrary time after the formation of the first hydrostatic core is also affecting the mass of the simulated core (Commer\c con \et\ 2011a).\\

Therefore, the different fragmentation level seen in our sample of massive dense cores could be a consequence of different initial magnetic field strength in the cores. For comparison, the magnetic field strength assumed in the $\mu=2$ case is around 0.3--1~mG at the scales of the massive dense core ($\sim3000$~AU), which is comparable to the magnetic field strength derived from observations of star-forming regions at similar scales, of about 0.5~mG (\eg\ Girart \et\ 2006; Falgarone \et\ 2008; Gon{\c c}alves \et\ 2008; Frau \et\ 2011). This indicates that the magnetic field strength required to inhibit significantly the fragmentation in a dense core for the simulations of Commer\c con \et\ (2011a) can easily be found in star-forming regions, and drives us to carry out further observational studies to study the relation of fragmentation and magnetic field strength in our sample.

\section{Conclusions}

We conducted 1.3~mm continuum and CO\,(2--1) PdBI observations, using its most extended configuration, towards four massive dense cores (I22172N, I22134, I22198, A5142), allowing us to study their fragmentation for mass sensitivities $\sim0.3$~\mo\ and spatial resolutions $\sim1000$~AU, comparable to studies of clusters in the optical/infrared. We complemented our high angular resolution observations with observations from the literature of 14 additional massive dense cores observed at similar mass sensitivities and spatial resolutions. For the 18 sources in the sample we estimated a set of parameters in a uniform way and assessed their role in the fragmentation process. Our main conclusions can be summarized as follows:

\begin{itemize}

\item[-] The four regions presented in this work show a variety of fragmentation levels. Towards I22172N we find one strong and elongated millimeter source, driving a small outflow, and surrounded by 3 fainter millimeter sources. The mass of the strongest millimeter source is estimated to be around 2--10~\mo, and its size is of $2300\times1000$~AU. I22134 reveals four low-mass $\lesssim1$~\mo\ fragments distributed surrounding the UCHII, for which we did not detect millimeter continuum emission. I22198 is the core with smallest degree of fragmentation, with one single source dominating the millimeter emission and no additional fragments down to masses of $\sim0.1$~\mo. The strong millimeter source, with a mass of about $\sim2$~\mo\ and a size of $500\times300$~AU, is driving at least one outflow. Finally, for the case of A5142 we find the highest degree of fragmentation. The millimeter emission consists of two strong elongated sources, both driving outflows, and with masses of about 5~\mo, which are surrounded by 5 additional unresolved fragments of about $\sim2$~\mo.

\item[-] For the total sample of 18 massive dense cores, we estimate luminosities in the range 300 to $2\times10^5$~\lo, single-dish masses of $\sim20$--2000~\mo, mean densities of (0.3--30)$\times10^5$~\cmt, CFE of 0.5--60\%, non-thermal velocity dispersions of 0.3--0.7~\kms, and rotational-to-gravitational energies of 0.003--0.072. 
For the 18 cores, the median number of milllimeter sources within a field of view of 0.1~pc is 3.25. 30\% of the cores show no signs of fragmentation, while $\sim50$\% split up into $\gtrsim4$ fragments. In addition, we assessed the number of infrared sources in the field of view, from 2MASS and Spitzer data available for the regions.

\item[-] We found no relation between the number of millimeter sources and any of the properties measured for the massive dense cores, and especially the correlation seen in the optical/infrared between \Lbol\ and the richness of a cluster is not seen in our sample of protoclusters.
Instead, there are hints for a correlation between the CFE and the density of the massive dense core,  which suggests that gravity dominates as density increases.
Another weak trend is found between the mass of the strongest millimeter source and the bolometric luminosity, 
suggesting a `simultaneous accretion picture', from core to fragments, and from fragments to protostars.
Furthermore, we found a correlation between the number of millimeter sources and the number of infrared+millimeter sources, indicating that probably the different episodes of star formation keep the same fragmentation level along time.

\item[-] Given the lack of correlation between the fragmentation level and the core properties in our sample, we compared our observations with the radiation magneto-hydrodynamic simulations of Commer\c con \et\ (2011a) of massive dense cores, and found that the low level of fragmentation seen in I22198 or I20126+4104 can be reproduced with their simulations for the highly magnetized core case (with $\mu=2$), while the high level of fragmentation seen in A5142 can be reproduced with a dense core with weak magnetization ($\mu=130$).

\end{itemize}

Overall, our observations suggest that the optical/infrared correlation previously found between the mass of the most massive star of the cluster and the cluster richness is not set at the early stages of cluster formation but probably arises as a result of cluster evolution. In addition, from this study we conclude that the fragmentation of massive dense cores, rather than being determined by the mass or density of the core, could be strongly influenced by its initial magnetic field.

%%%%%%%%%%%%%%%%%%%%%%%%%%%%%%%%%%%%%%%%%%%%%%%%%%%%%%%%%%%%%%%%%%%%%%%%%%%%%%%%
\acknowledgments
\begin{small}
The authors are grateful to the anonymous referee for valuable suggestions and comments which improved the quality of this paper.
A.P. is grateful to Javier Rod\'on, Marco Padovani, Pau Frau, Ray Furuya and James di Francesco for useful discussions, Herv\'e Bouy for help in the extraction of Spitzer/IRAC sources, and Javier Rod\'on, Floris van der Tak, Carlos Carrasco-Gonz\'alez, Silvia Leurini, Henrik Beuther, and Paula Teixeira for kindly providing their images for this work. 
A.P. is supported by a JAE-Doc CSIC fellowship co-funded with the European Social Fund under the program "Junta para la Ampliaci\'on de Estudios", by the Spanish MEC grant AYA2011-30228-C03-02 (co-funded with FEDER funds), and by the AGAUR grant 2009SGR1172 (Catalonia).
G.B. is funded by an Italian Space Agency (ASI) fellowship under contract number I/005/07/01.
The research of B.C. is supported by the postdoctoral fellowships from CNES and by the french ANR Retour Postdoc program.
This paper was partially supported by the Spanish MICINN program `CONSOLIDER INGENIO 2010: Molecular Astrophysics, Herschel-ALMA Era, ASTROMOL' (CSD2009-00038), and by the European Community's Seventh Framework Program (FP7/2007--2013) under agreement 229517. 
This research has made use of the SIMBAD database, operated at CDS, Strasbourg,  France, the NASA's Astrophysics Data System, Aladin, AKARI observations, a JAXA project with the participation of ESA; the Wide-field Infrared Survey Explorer, which is a joint project of the University of California, Los Angeles, and the Jet Propulsion Laboratory/California Institute of Technology, funded by NASA; the Two Micron All Sky Survey, which is a joint project of the University of Massachusetts and the Infrared Processing and Analysis Center/California Institute of Technology, funded by NASA and the National Science Foundation (NSF); the Midcourse Space Experiment, for which processing of the data was funded by the Ballistic Missile Defense Organization with additional support from NASA Office of Space Science; and the NASA/ IPAC Infrared Science Archive, which is operated by the Jet Propulsion Laboratory, California Institute of Technology, under contract with NASA.
\end{small}

%%%%%%%%%%%%%%%%%%%%%%%%%%%%%%%%%%%%%%%%%%%%%%%%%%%%%%%%%%%%%%%%%%%%%%%%%%%%%%%%

%\begin{appendix}

\appendix

\section{Complementary maps}

As outlined in Sect.~\ref{saprop}, we compiled a number of millimeter interferometric images published in the literature whose angular resolution and sensitivity are good enough to compare to the 4 regions studied in this work with the PdBI in A configuration. From these images we estimated the number of millimeter sources within 0.1~pc, $N_\mathrm{mm}$. In this Appendix we present the sample of 14 millimeter images (see Bontemps \et\ 2010 for the millimeter images of the four regions in Cygnus\,X).

\begin{figure*}[h]
\begin{center}
\begin{tabular}[b]{c}
    \epsfig{file=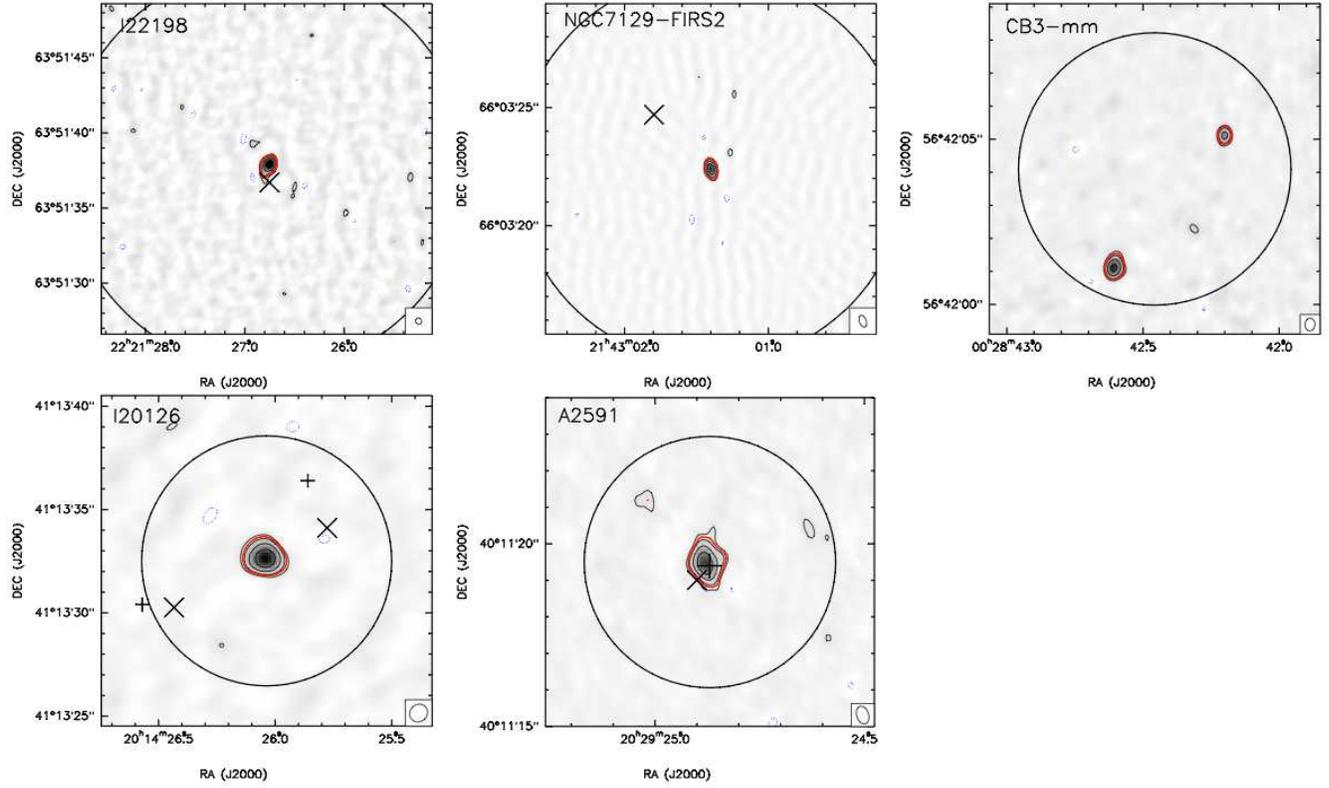, width=18cm, angle=0}\\
\end{tabular}
\caption{1.3~mm continuum high angular resolution maps for cores with 1--2 fragments. 
Contours for all regions are $-4$, 4, 8, 16, and 32 times the rms noise in each region, listed in Table~\ref{tstatistics}, along with references. Synthesized beams are plotted in the bottom-right corner of each panel, and the black circle corresponds to 0.1~pc diameter. Crosses correspond to Spitzer/IRAC sources at 3.6~\mum, and plus signs indicate the 2MASS sources. In all panels the red contour corresponds to the identification level of $6\sigma$.
}
\label{flit1}
\end{center}
\end{figure*}

\begin{figure*}[h]
\begin{center}
\begin{tabular}[b]{c}
    \epsfig{file=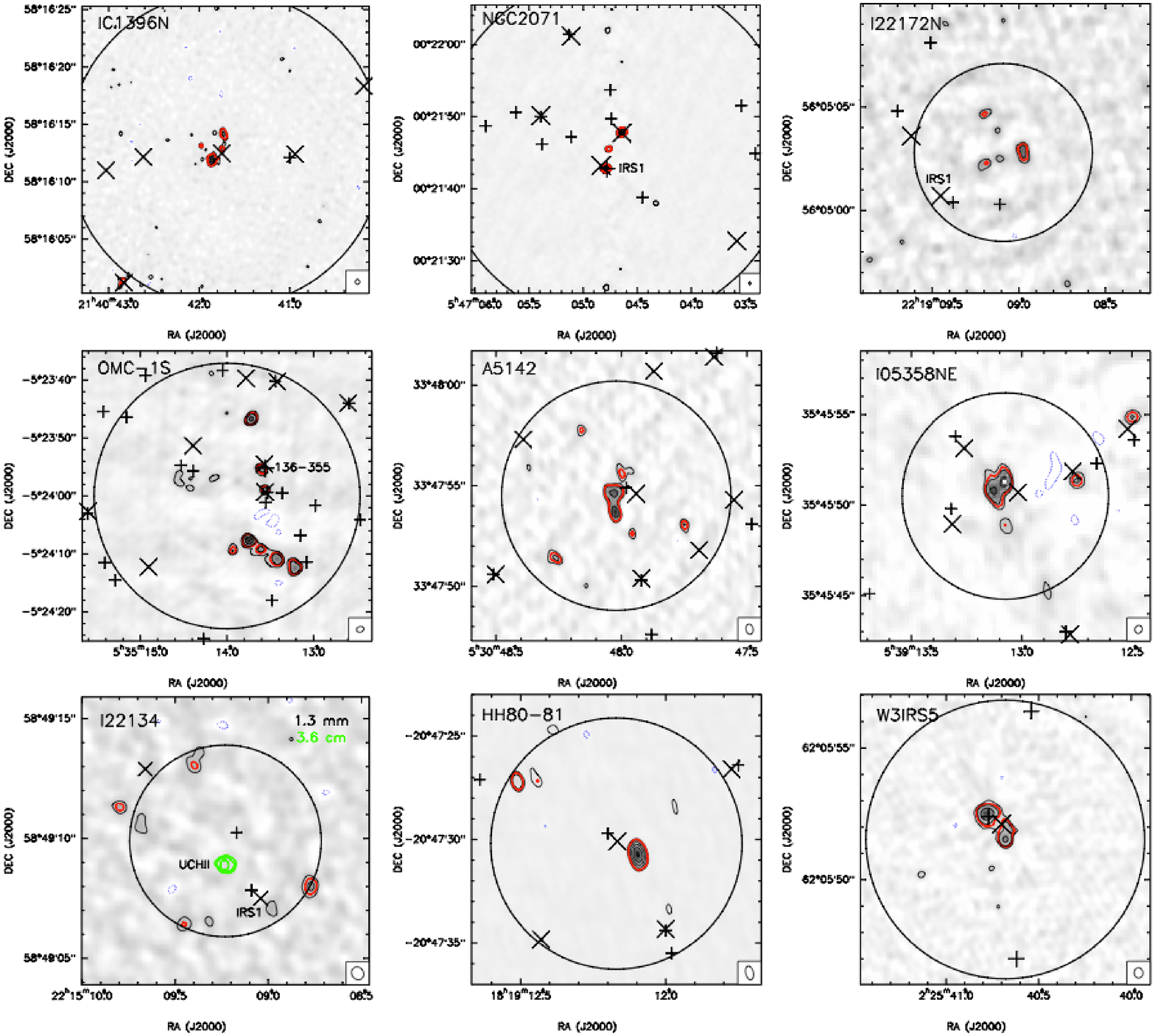, width=18cm, angle=0}
\end{tabular}
\caption{1.3~mm continuum high angular resolution maps for cores with $\ge3$ fragments (for NGC\,2071 it is the 3~mm image). 
Contours for all regions are $-4$, 4, 8, 16, 32, and 64 times the rms noise in each region. The rms noise used for each region is given in Table~\ref{tstatistics}, and references are given in the main text. Synthesized beams are plotted in the bottom-right corner of each panel. For I22134, we overplot in green contours the 3.6~cm emission (contours as in Fig.~\ref{fcont}). The symbols are the same as in Fig.~\ref{flit1} (red contour corresponds to the identification level of $6\sigma$). We labeled the source used for the SED in cases of possible confusion (if no source is labeled, we used the infrared source closest to the strongest millimeter source).
}
\label{flit2}
\end{center}
\end{figure*}

%\end{appendix}

\end{document}